\documentclass{article}

\usepackage{epsf}
\usepackage{rotate}

\begin{document}

\title{
\vskip-3cm{\baselineskip14pt
\centerline{\normalsize\hfill hep-ph/9512224}
\centerline{\normalsize\hfill TUM--HEP--227/95}
\centerline{\normalsize\hfill November 1995}}
\vskip0.6cm
The Goldstone boson equivalence theorem with fermions
\vskip1.0cm
}

\author{
Loyal Durand\thanks{Electronic address: ldurand@wishep.physics.wisc.edu}
                \\[0.4cm]
{\small Department of Physics, University of Wisconsin-Madison,}
                \\[-0.1cm]
{\small 1150 University Avenue, Madison, WI 53706}
                \\[0.8cm]
and
                \\[0.8cm]
Kurt Riesselmann\thanks{Electronic address: kurtr@physik.tu-muenchen.de}
                \\[0.4cm]
{\small Physik Department T30, Technische Universit\"{a}t M\"{u}nchen,}
                \\[-0.1cm]
{\small James-Franck Strasse, 85747 Garching b.\ M\"{u}nchen, Germany}
}
\date{}
\maketitle

\vskip0.6cm
\begin{abstract}
  The calculation of the leading electroweak corrections to physical transition
  matrix elements in powers of $M_H^2/v^2$ can be greatly simplified in the
  limit $M_H^2\gg M_W^2,\, M_Z^2$ through the use of the Goldstone boson
  equivalence theorem. This theorem allows the vector bosons $W^\pm$ and $Z$ to
  be replaced by the associated scalar Goldstone bosons $w^\pm$, $z$ which
  appear in the symmetry breaking sector of the Standard Model in the limit of
  vanishing gauge couplings.  In the present paper, we extend the equivalence
  theorem systematically to include the Yukawa interactions between the
  fermions and the Higgs and Goldstone bosons of the Standard Model. The
  corresponding Lagrangian ${\cal L}_{EQT}$ is given, and is formally
  renormalized to all orders. The renormalization conditions are formulated
  both to make connection with physical observables and to satisfy the
  requirements underlying the equivalence theorem.  As an application of this
  framework, we calculate the dominant radiative corrections to fermionic Higgs
  decays at one loop including the virtual effects of a heavy top quark.  We
  apply the result to the decays $H\rightarrow t\bar{t}$ and $H\rightarrow
  b\bar{b}$, and find that the equivalence theorem results including fermions
  are quite accurate numerically for Higgs-boson masses $M_H> 400\,(350)$ GeV,
  respectively, even for $m_t=175$ GeV.
\end{abstract}

\newpage

\section{Introduction}

The physics of the Higgs-sector is one of the most important subjects of
high-energy experiments in the near future.  To distinguish the Standard Model
(SM) Higgs-sector from other theoretical models, the calculation of radiative
corrections to high-energy processes may be very important.  In addition, the
effects of a Higgs boson on low-energy observables such as the $\rho$-parameter
already need to be included in present precision measurements. However, the
calculation of the complete higher-order corrections is a difficult task.  It
is therefore important if one can calculate the
most significant corrections relatively simply within a
well-defined approximation.  In the case of a heavy Higgs boson, such an
approximation is provided by the equivalence theorem (EQT) \cite{lee,chan}.
This theorem shows that the leading correction to a physical process
in powers of $M_H^2/v^2$ can be obtained at any order by replacing the vector
bosons $W^\pm$ and $Z$ of the SM by the associated
scalar Goldstone bosons $w^\pm$, $z$ which appear in the symmetry breaking
sector of the theory in the limit of vanishing gauge couplings.
This results in a substantial simplification of the calculations.
The equivalence theorem is known to hold for $M_H\gg M_W,\,M_Z$. An important
question, however, is how heavy the Higgs boson must be before the
equivalence theorem gives a good approximation to the full electroweak
theory.

The existence of a heavy top quark leads to a large Yukawa coupling $g_t$ in
the SM. In the case of a not-so-heavy Higgs boson, one can expect the radiative
corrections which involve $g_t$ to be comparable in magnitude to the
corrections which involve the quartic Higgs coupling $\lambda=M_H^2/2v^2$.  It
is therefore desirable to take these contributions into account, and to
calculate them in a simple way. This can again be done using the equivalence
theorem: in the limit of vanishing gauge couplings, the fermions couple only to
the Higgs boson and the Goldstone bosons in the symmetry breaking sector of the
theory.  However, the specific calculational scheme used has to be consistent
both with the EQT as formulated for the bosons \cite{lee,chan}, and with the
complete SM. In the present paper we show how this can be achieved by extending
the equivalence theorem, including fermions, and implementing consistent
renormalization conditions. The resulting framework is a foundation for future
calculations of leading higher-order electroweak corrections, and it is
applicable to all orders in $\lambda$ and $g_f$. Gauge interactions are
explicitly set to zero.

Previous calculations based on the EQT have mostly been concerned with
processes in which both the external and internal
particles are Higgs bosons $H$ or longitudinally
polarized gauge bosons $W_L$ and $Z_L$
\cite{lee,chan,passarino1,marciano,djl1,passarino2,djl2,djm,dmr,ghinculov2},
and have neglected corrections from internal fermion
loops. An exception is the work of Barbieri {\em et al.} \cite{barbieri},
who calculated the $m_t^4$ corrections to the decays $Z\rightarrow \mu^+\mu^-$
and $Z\rightarrow b\bar{b}$ using a method based on the equivalence
theorem, but with the fermion fields renormalized only to one loop.
The framework presented here allows the systematic inclusion of
internal fermion loops at higher orders, as well as the calculation of
radiative corrections to processes involving external fermions.

We illustrate our method at the one-loop level using the
fermionic decay of the Higgs boson, $H\rightarrow f\bar
f$. The equivalence theorem result approximates the full electroweak
correction to $\Gamma(H\rightarrow t\bar{t})$ very well for $M_H> 400$
GeV, but fails near the threshold at 350 GeV for $m_t=175$ GeV, where the
effects of the gauge interactions become important. The result for $\Gamma(
H\rightarrow b\bar{b})$ is also quite good for $M_H> 350$ GeV, though the
accuracy is reduced by a cancellation between the Yukawa and Higgs-boson
contributions. The gauge couplings give the dominant, very small, corrections
for lower values of $M_H$.

\section{The Lagrangian and renormalization scheme}

\subsection{Framework}

We will be concerned in later sections with the calculation of the leading
contributions to the decay rate of the Higgs boson to fermions, $H\rightarrow
f\bar{f}$, in the limit of large Higgs-boson mass, $M_H\gg M_W$. Our tool for
extracting the leading contributions in powers of $M_H/M_W$ will be the
Goldstone boson equivalence theorem \cite{lee,chan,corn,bagger,hvelt,he}.  This
theorem states that the dominant electroweak contribution to a Feynman graph
can be calculated by replacing the gauge bosons $W^\pm$, $Z$ of the full
electroweak theory by the would-be Goldstone bosons $w^\pm$, $z$ of the
symmetry-breaking sector of the theory, and ignoring the gauge couplings.  The
result is an expansion for the dominant contributions in powers of $G_FM_H^2$,
or equivalently, in powers of the quartic Higgs coupling $\lambda$.  In the
limit $g,\,g^\prime\rightarrow 0$, the tranverse gauge bosons decouple from the
fermions and the scalar fields, and can be ignored. The remainder of the
standard electroweak model reduces to a theory defined by the
equivalence-theorem Lagrangian ${\cal L}_{EQT}={\cal L}_H+{\cal L}_F$, where
${\cal L}_H$ is the Lagrangian for the scalar fields in the symmetry-breaking
sector, and ${\cal L}_F$ is the Lagrangian for the fermions which includes
their Yukawa interactions with the scalar fields.  This reduced theory gives a
good approximation to the full electroweak theory for $M_W^2\ll M_H^2$,
provided that the couplings are properly defined and the Goldstone modes are
renormalized at a momentum scale $p^2\ll M_H^2$ \cite{bagger}.

We will work entirely in the equivalence theorem limit using ${\cal L}_{EQT}$,
\begin{equation}
{\cal L}_{EQT}={\cal L}^0_H+{\cal L}^0_F+{\rm counterterms}, \label{lag}
\end{equation}
where
\begin{eqnarray}
{\cal L}^0_H &=&
{\textstyle\frac{1}{2}}\left(\partial_\mu\Phi\right)^\dagger
\left(\partial^\mu\Phi\right)-{\textstyle\frac{1}{4}}\lambda
\left(\Phi^\dagger\Phi\right)^2+
{\textstyle\frac{1}{2}}\mu^2\Phi^\dagger\Phi\,, \label{l0higgs}\\
{\cal L}^0_F &=&
i\bar{\Psi}_L\!\not\!\partial\,\Psi_L+i\bar{\psi}_{t,R}\!\not\!\partial
\,\psi_{t,R}+i\bar{\psi}_{b,R}\!\not\!\partial\,\psi_{b,R} \nonumber\\
&&\!-{\textstyle\frac{1}{\!\!\sqrt{2}}}g_t\bar{\Psi}_L\,\tilde{\Phi}\,
\psi_{t,R}-{\textstyle\frac{1}{\!\!\sqrt{2}}}g_b\bar{\Psi}_L\,\Phi\,
\psi_{b,R}+{\rm h.c.}+\cdots\,. \label{l0fermi}
\end{eqnarray}
We have written only the top- and bottom quark contributions to ${\cal L}^0_F$;
the remaining fermionic contributions have the same form, with no righthanded
contributions for the neutrinos.  In these expressions, $\Psi_L$, $\Phi$, and
$\tilde{\Phi}$ are SU(2)$_L$ doublets with normalizations defined by
\begin{eqnarray}
\Psi_L &=& \left(
\begin{array}{c}
        \psi_{t,L}\\
        \psi_{b,L}
\end{array}
\right),\qquad
\psi_{f,L}={\textstyle\frac{1}{2}}(1-\gamma^5)\,\psi_f,\label{psil}\\
\Phi &=& \left(
\begin{array}{c}
        w_1+iw_2\\
        h+iz
\end{array}
\right)=\left(
\begin{array}{c}
        \sqrt{2}w^{+}\\
        h+iz
\end{array}
\right), \label{phi} \\
\tilde{\Phi} &=& i\sigma_2\Phi^*=\left(
\begin{array}{c}
        h-iz\\
        -w_1+iw_2
\end{array}
\right)=\left(
\begin{array}{c}
        h-iz\\
        -\sqrt{2}w^{-}
\end{array}
\right)\,. \label{phitilde}
\end{eqnarray}
The righthanded fields are SU(2)$_L$ singlets, with
\begin{equation}
\psi_{f,R}={\textstyle\frac{1}{2}}(1+\gamma^5)\,\psi_f\,. \label{psir}
\end{equation}

The expressions in Eqs.\ (\ref{l0higgs}) and (\ref{l0fermi}) include all
possible SU(2)$_L\times$U(1)$_Y$ symmetric terms consistent with
renormalizability of the Lagrangian. The counterterms necessary to effect the
renormalization without breaking the symmetry must have the same forms, and can
be introduced by independent multiplicative rescalings of each term above.
Thus, in the case of the Higgs Lagrangian, all possible symmetric counterterms
are generated by multiplying the kinetic energy term by a factor $Z_\Phi$,
replacing $\lambda$ by $\lambda+\delta\lambda$, and $\mu^2$ by
$\mu^2+\delta\mu^2$. Because the minimum in the Higgs potential is at a nonzero
value of $\Phi^\dagger\Phi$ for $\mu^2>0$, $\Phi$ has a nonzero expectation
value in the physical vacuum,
\begin{equation}
\langle\,\Omega\,|\,\Phi^\dagger\Phi\,|\,\Omega\,\rangle=v^2\,. \label{vev}
\end{equation}
We will absorb the vacuum expectation value by rewriting the field $h$ as
$h\rightarrow H+v$. This results in an expansion around the physical vacuum
in which
\begin{equation}
\langle\,\Omega\,|\,H\,|\,\Omega\,\rangle=\langle\,\Omega\,|\,{\bf w}\,
|\,\Omega\,\rangle=0\,.
\label{vac}
\end{equation}
The renormalized Higgs Lagrangian then has the form
\begin{eqnarray}
{\cal L}_H &=& {\textstyle\frac{1}{2}}Z_\Phi\left(
\partial_\mu{\bf w}\cdot\partial^\mu{\bf w}
+\partial_\mu H\,\partial^\mu H\right)
-{\textstyle\frac{1}{4}}\left(\lambda+\delta\lambda\right)\left({\bf w}^2
+H^2+2vH\right)^2 \nonumber\\
&&-{\textstyle\frac{1}{2}}\left(\left(\lambda+\delta\lambda
\right)v^2-\mu^2-\delta\mu^2\right)\left({\bf w}^2+H^2+2vH\right)\,,
\label{l1higgs}
\end{eqnarray}
where $\bf w$ is the SO(3) vector $(w_1,\,w_2,\,w_3)$ with $w_3=z$, and
we have dropped an additive constant.

It is convenient for calculation to define a set of rescaled or
``bare'' fields $H_0$, ${\bf w}_0$ and the corresponding SU(2)$_L$ doublet
$\Phi_0$ such that the kinetic terms in ${\cal L}_H$
have the customary unit normalization,
\begin{equation}
H_0=Z_\Phi^{1/2}H\,,\qquad {\bf w}_0=Z_\Phi^{1/2}{\bf w}\,,
\qquad \Phi_0=Z_\Phi^{1/2}\Phi\,,\label{barehiggs}
\end{equation}
and to introduce a corresponding bare vacuum expectation value
$v_0$, a bare coupling $\lambda_0$, and a parameter $\delta m_0^2$,
\begin{eqnarray}
v_0 &=& Z_\Phi^{1/2}v, \label{v0}\\
\lambda_0 &=& \frac{\lambda}{Z_\Phi^2}\left(1+\frac{\delta\lambda}{\lambda}
\right)\,,\label{lambda0}\\
\delta m_0^2 &=& Z_\Phi^{-1}\left(\mu^2+\delta\mu^2-\left(\lambda
+\delta\lambda\right)v^2\right)\,.\label{dm2}
\end{eqnarray}
With this rescaling, we obtain the form of ${\cal L}_H$ that we will use,
\begin{eqnarray}
{\cal L}_H &=& {\textstyle\frac{1}{2}}\left(\partial_\mu{\bf w_0}\cdot
\partial^\mu{\bf w_0}+\partial_\mu H_0\,\partial^\mu H_0\right)
-{\textstyle\frac{1}{4}}
\lambda_0\left({\bf w}_0^2+H_0^2+2v_0H_0\right)^2 \nonumber\\
&& +{\textstyle\frac{1}{2}}
\delta m_0^2\left({\bf w}_0^2+H_0^2+2v_0H_0\right)\,.\label{lhiggs}
\end{eqnarray}

The fermion Lagrangian ${\cal L}_F$ can be treated in a similar fashion.  There
is a separate SU(2)$_L\times$U(1)$_Y$-symmetric counterterm for each term in
Eq.\ (\ref{l0fermi}). These counterterms can be absorbed in the definitions of
a set of bare fields and couplings to bring ${\cal L}_F$ into the form
\begin{eqnarray}
{\cal L}_F &=& i\bar{\Psi}_L^0\!\not\!\partial\,\Psi_L^0
+i\bar{\psi}_{t,R}^0\!\not\!\partial\,\psi_{t,R}^0
+i\bar{\psi}_{b,R}^0\!\not\!\partial\,\psi_{b,R}^0 \nonumber\\
&&\!-\textstyle{\frac{1}{\!\!\sqrt{2}}}g_t^0\,\bar{\Psi}_L^0
\,\tilde{\Phi}_0\,\psi_{t,R}^0-\textstyle{\frac{1}{\!\!\sqrt{2}}}
g_b^0\,\bar{\Psi}_L^0
\,\Phi_0\,\psi_{b,R}^0+{\rm h.c.}\,,\label{lfermi}
\end{eqnarray}
where
\begin{eqnarray}
\Psi^0_L &=& Z_L^{1/2}\,\Psi_L\,,\label{zleft}\\
\psi^0_{f,R} &=& Z_{f,R}^{1/2}\,\psi_{f,R}\,,\quad f=t,\,b,\label{zfright}\\
g_f^0 &=& \frac{g_f}{Z_\Phi^{1/2}}\left(1+\frac{\delta g_f}{g_f}\right)\,.
\label{deltagf}
\end{eqnarray}
Note that there are separate renormalization constants for the $\psi_{t,R}$
and $\psi_{b,R}$, and a single renormalization constant for the left-handed
doublet $\Psi_L$ \cite{boehm}. The factor $Z_\Phi^{-1/2}$ in $g_f^0$ has been
introduced for later convenience.

\subsection{Fixing the renormalization of the Higgs Lagrangian}
\label{sec.lfrenorm}

In order that the theory defined by ${\cal L}_{EQT}$ correspond to the
equivalence theorem limit of the standard electroweak model, the definitions of
the couplings and the renormalization scheme must be chosen to be consistent
with that limit. Such a choice is not automatic. It is necessary that: ({\em
  i\/}), the $w^\pm$ and $z$ fields be renormalized at a momentum scale $p^2$
with $|p^2|\ll M_H^2$ \cite{bagger}; and ({\em ii\/}), the couplings be defined
in terms of physical standard-model quantities which have well-defined limits
for $g,g'\rightarrow 0$.  In the following, we will use an on-mass-shell
renormalization scheme, define the quartic Higgs coupling in terms of the Fermi
constant $G_F$ and the physical Higgs-boson mass $M_H$, and relate the Yukawa
couplings to the physical masses of the fermions.

The $w^\pm$ and $z$ bosons are guaranteed to be massless by the Goldstone
theorem \cite{goldstone}.  We will therefore renormalize the $w_0^\pm$ and
$z_0$ fields at $p^2=0$, thus satisfying condition ({\em i\/}), and will
renormalize the Higgs field at $p^2=M_H^2$.  The quartic coupling $\lambda$
will be defined \cite{sirlin} to be given {\em exactly} by the relation
\begin{equation}
\lambda=M_H^2/2v^2=G_FM_H^2/\sqrt{2},\label{lambda}
\end{equation}
where $G_F$ is the Fermi constant obtained from the muon decay rate
using the standard electromagnetic radiative corrections, and
$v$ is the physical vacuum expectation value, $v=2^{-1/4}G_F^{-1/2}$.
While this definition involves a process at a low momentum transfer, it
connects smoothly with equivalence theorem limit as shown in
\cite{sirlin,maher}.

The determination of the renormalization constants proceeds as follows.  The
real part of the two-point function or inverse propagator $\Gamma^{(2)}(p^2)$
for each of the particles must vanish for $p^2$ equal to the square of the
physical mass of that particle, $\Gamma^{(2)}(m^2)=0$. The two-point functions
calculated using the bare fields are easily seen to have the form
\begin{eqnarray}
\Gamma_{w_0}^{(2)}(p^2)&=&p^2-\Pi^0_w(p^2)+\delta m_0^2\,,\label{gammaw}\\
\Gamma_{z_0}^{(2)}(p^2)&=&p^2-\Pi^0_z(p^2)+\delta m_0^2\,,\label{gammaz}\\
\Gamma_{H_0}^{(2)}(p^2)&=&p^2-\Pi_H^0(p^2)-2\lambda_0 v_0^2+\delta m_0^2\,,
\label{gammah}
\end{eqnarray}
where the $\Pi$'s are the bare self-energy functions.  Since there is only a
single mass counterterm $\delta m_0^2$ in the Higgs Lagrangian, Eq.\
(\ref{lhiggs}), the vanishing of the renormalized masses $m_w$ and $m_z$
required by the Goldstone theorem leads to the relations
\begin{equation}
\delta m_0^2=\Pi^0_w(0)=\Pi^0_z(0)\,,\label{pi(0)}
\end{equation}
and determines $\delta m_0^2$. The self-energy functions for the bare $w^\pm$
and $z$ fields must therefore be equal at $p^2=0$, an identity that holds to
all orders in perturbation theory. This would not be surprising in the absence
of Yukawa couplings since the Goldstone fields then have an SO(3) symmetry, and
$\Pi^0_w(p^2)$ and $\Pi^0_z(p^2)$ are necessarily identical for all values of
$p^2$. However, the presence of fermions with unequal Yukawa couplings in
${\cal L}_{EQT}$ breaks the SO(3) symmetry, and destroys the equality of the
self-energy functions except at the special point $p^2=0$ \cite{lytel}.  The
equality $\Pi_w^0(0)=\Pi_z^0(0)$ can be checked to one loop using the results
we give later.  In the following, we will replace $\delta m_0^2$ by
$\Pi^0_w(0)$.

The bare coupling $\lambda_0$ can be determined by using the definition
$\lambda=M_H^2/2 v^2$ and Eq.\ (\ref{gammah}). The on-mass-shell
condition for the Higgs boson, $\Gamma_{H_0}^{(2)}(M_H^2)=0$, gives
the relation
\begin{equation}
M_H^2=2\lambda_0 v_0 + \Pi_H^0(M_H^2) - \delta m_0^2.\label{massshell}
\end{equation}
Upon replacing $v_0^2$ by $Z_\Phi v^2=Z_\Phi M_H^2/2\lambda$ and
$\delta m_0^2$ by $\Pi_w^0(0)$ in this expression and solving for
$\lambda_0$, we find that \cite{ReMH}
\begin{equation}
\lambda_0=\lambda\left(1+\frac{\delta\lambda}{\lambda}\right)\frac{1}
{Z_\Phi^2}
=\frac{\lambda}{Z_\Phi}\left(1-\frac{{\rm Re}\Pi^0_H(M_H^2)-\Pi^0_w(0)}
{M_H^2}
\right)\,.\label{l0}
\end{equation}

\begin{figure}[t]
\vspace*{33pt}
\centerline{
\epsfysize=1.6in {\epsffile{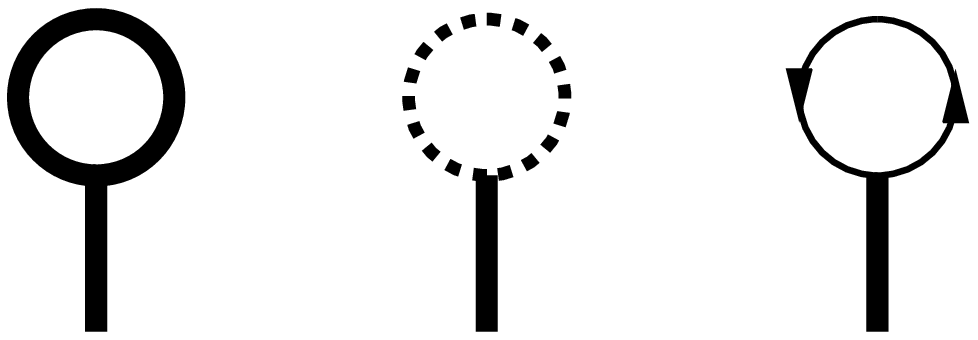}}
}
\vspace{1.1in}
\caption{The one-loop tadpole diagrams  which are cancelled by a counterterm to
  avoid a shift in the vacuum expectation value of the Higgs field.  Thick
  lines correspond to the Higgs particle, dotted lines represent the massless
  Goldstone bosons, and solid lines with arrows refer to fermions.
  A summation over all Goldstone boson and fermion loops is implied.  }
\label{figtadpoles}
\end{figure}

The condition that $v$ be the physical vacuum expectation value requires the
vanishing of the truncated one-point function for the Higgs field,
\begin{equation}
\Gamma_{H_0}^{(1)}(0)=-iT_0+iv_0\delta m_0^2=0,\label{tadpole}
\end{equation}
where $T_0$ is the sum of all Higgs tadpole graphs (see Fig.\
\ref{figtadpoles}) calculated using the bare fields.
This gives the further relation $\delta m_0^2=T_0/v_0$, so $\delta m_0^2$ can
be calculated either as a self energy or as a tadpole contribution.  The
resulting identity,
\begin{equation}
\Pi_w^0(0)=T_0/v_0,\label{T/v=Pi}
\end{equation}
provides a useful check on the calculations \cite{tadcalc}.  The vanishing of
the one-point function, Eq.\ (\ref{tadpole}), implies that the tadpole diagrams
and the tadpole counterterm in Eq.\ (\ref{lhiggs}) cancel order-by-order in the
perturbation expansion and can be dropped together, as discussed by Taylor
\cite{taylor}. We will adopt this simplification in the following calculations.

Finally, the wave function renormalization constants $Z_w$, $Z_z$, and $Z_H$
which relate the bare fields $w_0^\pm$, $z_0$, $H_0$ to physical fields,
\begin{equation}
w_0^\pm=Z_w^{1/2}w_{\rm phys}^\pm,\qquad z_0=Z_z^{1/2}z_{\rm phys},
\qquad H_0=Z_H^{1/2}H_{\rm phys}, \label{wphys}
\end{equation}
are determined by the condition that the propagators for the physical fields
have unit residue at the particle poles. The bare propagators do not, but
instead have residues $Z_i$ given by
\begin{equation}
Z_i\;=\;\left(\left.\frac{d}{dp^2}\Gamma_i^{(2)}(p^2)\right|_{p^2=m_i^2}
\right)^{-1}\,.\label{zdef}
\end{equation}
Thus, using Eqs.\ (\ref{gammaw}-\ref{gammah}),
\begin{eqnarray}
\frac{1}{Z_w}&=&\left.1-\frac{d}{dp^2}\Pi^0_w(p^2)\right|_{p^2=0},
\label{zw}\\
\frac{1}{Z_z}&=&\left.1-\frac{d}{dp^2}\Pi^0_z(p^2)\right|_{p^2=0},
\label{zz}\\
\frac{1}{Z_H}&=&\left.1-\frac{d}{dp^2}\Pi^0_H(p^2)\right|_{p^2=M_H^2}\,.
\label{zh}
\end{eqnarray}
In the presence of fermions with different Yukawa couplings or masses, e.g.,
top and bottom quarks with $m_t\not=m_b$, the self-energy functions $\Pi^0_w$
and $\Pi^0_z$ are not equal except at the isolated point $p^2=0$, and, as a
result, $Z_z\not=Z_w$.

The single wave function renormalization constant $Z_\Phi$
introduced earlier is sufficient, along with the other renormalization
constants, to absorb the divergences in the Higgs sector of the theory.
The constants $Z_w$, $Z_z$, and $Z_H$ can therefore differ from
$Z_\Phi$ only by finite multiplicative factors $\tilde{Z}_i$,
\begin{equation}
Z_i=\tilde{Z}_iZ_\Phi\,.\label{ztilde}
\end{equation}

The problem which remains is that of determining $Z_\Phi$, an essential step in
connecting the scalar theory to the full electroweak theory through the
equivalence theorem.  The crucial observation is that the Fermi constant $G_F$,
and therefore the vacuum expectation value $v$, is defined through
charged-current processes, i.e., muon decay and superallowed nuclear beta
decays. These involve the $W$ rather than the $Z$ boson.  The connection can
then be made using the Ward identities for the electroweak charged current
which underlie the equivalence theorem \cite{bagger,barbieri}.  In particular,
Bagger and Schmidt \cite{bagger} show that $W^\pm$ scattering amplitudes
calculated in the full electroweak theory are related to those calculated using
the equivalence theorem by powers of the ratio \cite{me}
\begin{equation}
C=\frac{M_W^0}{M_W}\frac{Z_W^{1/2}}{Z_w^{1/2}}=\frac{g_0v_0}{gv}
\frac{Z_W^{1/2}}{Z_w^{1/2}}=\frac{Z_\Phi^{1/2}}{Z_w^{1/2}}\left(1+O(g^2)
\right)\,.\label{C}
\end{equation}
They then use the Ward identity to establish that $C=1$ up to corrections of
order $g^2$ under conditions satisfied by our renormalization scheme. Thus,
$Z_\Phi=Z_w$ and $\tilde{Z}_w=1$ in the limit $g,g'\rightarrow 0$.

\subsection{Fixing the renormalization of the fermion Lagrangian}
\label{sec.fermrenorm}

When written with the SU(2)$_L$ spinor products expanded, the fermion
Lagrangian ${\cal L}_F$ in Eq.\ (\ref{lfermi}) assumes the form
\begin{eqnarray}
{\cal L}_F&=&i\bar{\psi}_t^0\!\not\!\partial\,\psi_t^0-(m_t+\delta m_t)
\bar{\psi}_t^0\,\psi_t^0-
i\bar{\psi}_b^0\!\not\!\partial\,\psi_b^0-(m_b+\delta m_b)
\bar{\psi}_b^0\,\psi_b^0 \nonumber\\
&&-{\textstyle\frac{1}{\sqrt{2}}}g_t^0\,\bar{\psi}_t^0\,H_0\,\psi_t^0
+{\textstyle \frac{i}{\sqrt{2}}}g_t^0\,\bar{\psi}_t^0\,\gamma^5z_0\,
\psi_t^0
-{\textstyle\frac{1}{\sqrt{2}}}g_b^0\,\bar{\psi}_b^0\,H_0\,\psi_b^0
-{\textstyle\frac{i}{\sqrt{2}}}g_b^0\,\bar{\psi}_b^0\,\gamma^5z_0\,\psi_b^0
\nonumber\\
&&+g_t^0\,\bar{\psi}_{t,R}^0\,w_0^+\,\psi_{b,L}^0
+g_t^0\,\bar{\psi}_{b,L}^0\,w_0^-\,\psi_{t,R}^0
-g_b^0\,\bar{\psi}_{b,R}^0\,w_0^+\,\psi_{t,L}^0
-g_b^0\,\bar{\psi}_{t,L}^0\,w_0^-\,\psi_{b,R}^0\,, \label{lf}
\end{eqnarray}
where we have used the relations in Eqs.\ (\ref{barehiggs}), (\ref{v0}), and
(\ref{deltagf}). The parameters $m_f$ and $g_f$ are defined to be the physical
masses and Yukawa couplings of the quarks while $\delta m_f$ is the mass or
coupling counterterm,
\begin{equation}
m_f=g_f\frac{v}{\sqrt{2}},\qquad \delta m_f=m_f\frac{\delta g_f}{g_f}.
\label{m,dm}
\end{equation}

Our calculations will be carried out using the bare fields and the standard
free propagators for the quarks, and yield corrected fermion propagators
$iS_F^0$ of the form
\begin{equation}
iS_F^0=\frac{i}{\not\! p-m_f-\Sigma^0_f-\delta m}\,. \label{s0f}
\end{equation}
The self-energy $\Sigma^0_f$ contains an axial vector component as
well as the vector and scalar components familiar in parity-conserving
theories. In particular, suppressing the quark label,
\begin{equation}
\Sigma^0=\not\! p\,\Pi^0_V(p^2)+\not\! p\gamma^5\Pi^0_A(p^2)+m\Pi^0_S(p^2)\,.
\label{sigma0}
\end{equation}
The physical propagators $iS_F$ are related to the bare propagators
through the connection between the physical and bare fields,
\begin{equation}
\psi=\psi_R+\psi_L=\left(Z_R^{-1/2}P_R
+Z_L^{-1/2}P_L\right)\psi^0, \label{psi}
\end{equation}
where $P_R=\frac{1}{2}(1+\gamma^5)$ and $P_L=\frac{1}{2}(1-\gamma^5)$.
Recalling the definition of the propagator, $iS_F(x-y)=$
$\langle\,\Omega\,|T\left(\psi(x),\bar{\psi}(y)\right)|\,\Omega\,\rangle$, we
find that
\begin{equation}
iS_F(p)=\left(Z_R^{-1/2}P_R+Z_L^{-1/2}P_L\right)iS_F^0(p)
\left(Z_R^{-1/2}P_L+Z_L^{-1/2}P_R\right)\,,\label{sfdef}
\end{equation}
or, calculating the inverse of $S^0_F$ explicitly,
\begin{eqnarray}
iS_F(p)&=&\left\{\not\! p\left[{\textstyle\frac{1}{2}}
\left(Z_L^{-1}+Z_R^{-1}\right)
\left(1-\Pi^0_V(p^2)\right)-{\textstyle\frac{1}{2}}
\left(Z_L^{-1}-Z_R^{-1}\right)
\Pi_A^0(p^2)\right]\right. \nonumber\\
&&\qquad +\not\!p\gamma^5\left[{\textstyle\frac{1}{2}}\left(Z_L^{-1}-
Z_R^{-1}\right)\left(
1-\Pi^0_V(p^2)\right)-{\textstyle\frac{1}{2}}
\left(Z_L^{-1}+Z_R^{-1}\right)\Pi^0_A(p^2)
\right]\nonumber\\
&&\qquad\qquad\left.+Z_L^{-1/2}Z_R^{-1/2}\left(m +\delta m
+m\,\Pi_S^0(p^2)
\right)\right\}\nonumber\\
&&\times\left\{p^2\left[\left(1-\Pi^0_V(p^2)\right)^2-\Pi_A^{0\,2}(p^2)
\right]-\left(m+\delta m+m\Pi_S^0\right)^2\right\}^{-1}\,.
\label{sf}
\end{eqnarray}

The mass or coupling counterterm $\delta m=m\,\delta g/g$ and the wave function
renormalization constants $Z_R$ and $Z_L$ are determined by the condition that
the physical propagator describe a freely propagating particle with a mass $m$
and unit residue at the particle pole, $S_F\rightarrow$
$1/(\not\!p-m)=(\not\!p+m)/(p^2-m^2)$ for $\not\!p\rightarrow m$.  $S_F$ has a
simple pole at $p^2=m^2$ provided that \cite{fermionmass}
\begin{equation}
1+\frac{\delta m}{m}=1+\frac{\delta g}{g}=
\sqrt{\left[1-\Pi^0_V(m^2)\right]^2-\Pi_A^{0\,2}(m^2)}
\ -\ \Pi^0_S(m^2). \label{deltag/g}
\end{equation}
The coefficient of $\not\!p\gamma^5$ in Eq.\ (\ref{sf}) must vanish if $S_F$ is
to be a normal massive propagator with equal right- and left-handed residues at
the particle pole.  This requires that
\begin{equation}
\left(Z_R-Z_L\right)\left[1-\Pi^0_V(m^2)\right]=\left(Z_R+Z_L\right)
\Pi^0_A(m^2). \label{condition1}
\end{equation}
The coefficients of $\not\!p$ and $m$ in the numerator will be equal if,
in addition,
\begin{equation}
{\textstyle\frac{1}{2}}\left(Z_R+Z_L\right)\left[1-\Pi^0_V(m^2)\right]
-{\textstyle\frac{1}{2}}\left(Z_R-Z_L\right)\Pi^0_A(m^2)=Z_R^{1/2}Z_L^{1/2}
\left[1+\frac{\delta g}{g}+\Pi^0_S(m^2)\right]. \label{condition2}
\end{equation}

Equations (\ref{condition1}) and (\ref{condition2}) are homogeneous of degree
one in the $Z$'s, and only determine the ratio $Z_R/Z_L$. The two equations may
also be combined to obtain the pole condition, Eq.\ (\ref{deltag/g}).  The
magnitude of the $Z$'s is determined by the final condition, that $S_F$ have
unit residue at the pole. With the ratios $Z_R/Z_L$ determined as above for
both the quarks in a doublet, this final condition can only be enforced for one
of the two quark propagators by adjusting the single remaining renormalization
constant $Z_L$ for the left handed fields $\Psi_L$, Eq.\ (\ref{zleft}). The
second quark field requires an additional wave function renormalization to
reach the standard normalization. This extra renormalization constant is
finite; the infinities in the wave function renormalizations can be be absorbed
entirely in $Z_L$ and the two $Z_R$'s, which generate the only symmetric
counterterms allowed in the kinetic part of ${\cal L}_F$, Eq.\ (\ref{lfermi}).
Since our calculations are done entirely in terms of the bare fields, with the
$Z$'s appearing only in the overall factors necessary for external particles,
the choice of definition for the original $Z_L$ is irrelevant, and we will give
only the final renormalization constants. These are determined by setting the
ratio of the coefficient of $\not\!p$ in Eq.\ (\ref{sf}) to the derivative of
the denominator equal to unity for $p^2=m^2$. Using Eqs.\
(\ref{deltag/g}-\ref{condition2}), we find that
\begin{eqnarray}
\frac{1}{Z_L}&=&1-\Pi^0_V(m^2)+\Pi^0_A(m^2)-2m^2\Pi^{0\;\prime}_S(m^2)
\sqrt{\frac{1-\Pi^0_V(m^2)+\Pi^0_A(m^2)}{1-\Pi^0_V(m^2)-\Pi^0_A(m^2)}}
\nonumber\\
&&\qquad -2m^2\frac{\left(1-\Pi^0_V(m^2)\right)\Pi^{0\;\prime}_V(m^2)
-\Pi^0_A(m^2)\Pi^{0\;\prime}_A(m^2)}{1-\Pi^0_V(m^2)-\Pi^0_A(m^2)}\label{zl}
\end{eqnarray}
and
\begin{eqnarray}
\frac{1}{Z_R}&=&1-\Pi^0_V(m^2)-\Pi^0_A(m^2)-2m^2\Pi^{0\;\prime}_S(m^2)
\sqrt{\frac{1-\Pi^0_V(m^2)-\Pi^0_A(m^2)}{1-\Pi^0_V(m^2)+\Pi^0_A(m^2)}}
\nonumber\\
&&\qquad -2m^2\frac{\left(1-\Pi^0_V(m^2)\right)\Pi^{0\;\prime}_V(m^2)
-\Pi^0_A(m^2)\Pi^{0\;\prime}_A(m^2)}{1-\Pi^0_V(m^2)-\Pi^0_A(m^2)}\,,\label{zr}
\end{eqnarray}
where $\Pi'(p^2)=d\Pi(p^2)/dp^2$.  These expressions hold to all orders in
perturbation theory, with one set for each quark. At the one-loop level
\cite{boehm} of interest here, they simplify substantially to
\begin{eqnarray}
Z_L&=&1
+\Pi^0_V(m^2)-\Pi^0_A(m^2)+2m^2\Pi^{0\;\prime}_V(m^2)
+2m^2\Pi^{0\;\prime}_S(m^2),\label{zl1}\\
Z_R&=&1
+\Pi^0_V(m^2)+\Pi^0_A(m^2)+2m^2\Pi^{0\;\prime}_V(m^2)
+2m^2\Pi^{0\;\prime}_S(m^2),\label{zr1}\\
\frac{\delta g}{g}&=&\frac{\delta m}{m}=-\Pi^0_V(m^2)-\Pi^0_S(m^2). \label{dg1}
\end{eqnarray}

It is straightforward, finally, to establish the renormalization factors which
must be used for the external fermions when scattering amplitudes are
calculated using the bare fields. By using the standard reduction formulas
\cite{bjorken}, we can express the physical scattering amplitudes in terms of
Fourier transforms of vacuum expectation values of time-ordered products of the
physical fields. These vacuum expectation values appear multiplied by a factor
$(\not\!p-m)$ and a spinor for each ingoing or outgoing fermion line.  The
physical fermion fields $\psi$, $\bar{\psi}$ in the time-ordered products can
be replaced by the bare fields $\psi^0$, $\bar{\psi}^0$ using the definition in
Eq.\ (\ref{psi}).  When the result is reexpressed in terms of the bare
truncated Green's function $\Gamma^0_n$, the external factors are replaced by
$S_F^0(p)(Z_R^{-1/2}P_L+Z_L^{-1/2}P_R)(\not\!p-m)$ and a spinor for each
ingoing fermion line, and by $(\not\!p-m)(Z_R^{-1/2}P_R+Z_L^{-1/2}P_L)S_F^0(p)$
and a conjugate spinor for each outgoing line.  However, it follows from the
first line of Eq.\ (\ref{sf}) and the normalization of the physical propagator
that
\begin{eqnarray}
S_F^0(p)(Z_R^{-1/2}P_L+Z_L^{-1/2}P_R) &=& (Z_R^{1/2}P_R+Z_L^{1/2}
P_L)S_F(p),\\
(Z_R^{-1/2}P_R+Z_L^{-1/2}P_L)S_F^0(p) &=&  S_F(p)(Z_R^{1/2}P_L
+Z_L^{1/2}P_R) ,\label{lines}
\end{eqnarray}
and that
\begin{equation}
S_F(p)(\not\!p-m)=(\not\!p-m)S_F(p)=1,\quad \not\!p=m.
\end{equation}
As a result, the physical scattering amplitudes are given by the
truncated Green's functions $\Gamma^0_n$ calculated in terms of the bare
fields, multiplied by a factor $(Z_R^{1/2}P_R+Z_L^{1/2}P_L)$
and the appropriate spinor for each incoming fermion line,
and by a factor $(Z_R^{1/2}P_L+Z_L^{1/2}P_R)$ and a conjugate
spinor for each outgoing fermion line. These factors generalize
the usual factors of $Z^{1/2}$ for standard Dirac fields to the case
of chiral interactions.

\section{ $H\rightarrow f\bar f$ to one loop}

In the present section, we will sketch the calculation of the one-loop
corrections to the matrix element for the decay $H\rightarrow f\bar f$ using
the equivalence theorem, and compare this approximate result, valid for $M_H\gg
M_W$ independently of the fermion masses, with the exact result obtained by
other authors \cite{kniehl,hollik,bardin,hff}.

\subsection{Form of the decay matrix element}

\begin{figure}[t]
\vspace*{33pt}
\centerline{
\epsfysize=3.7in {\epsffile{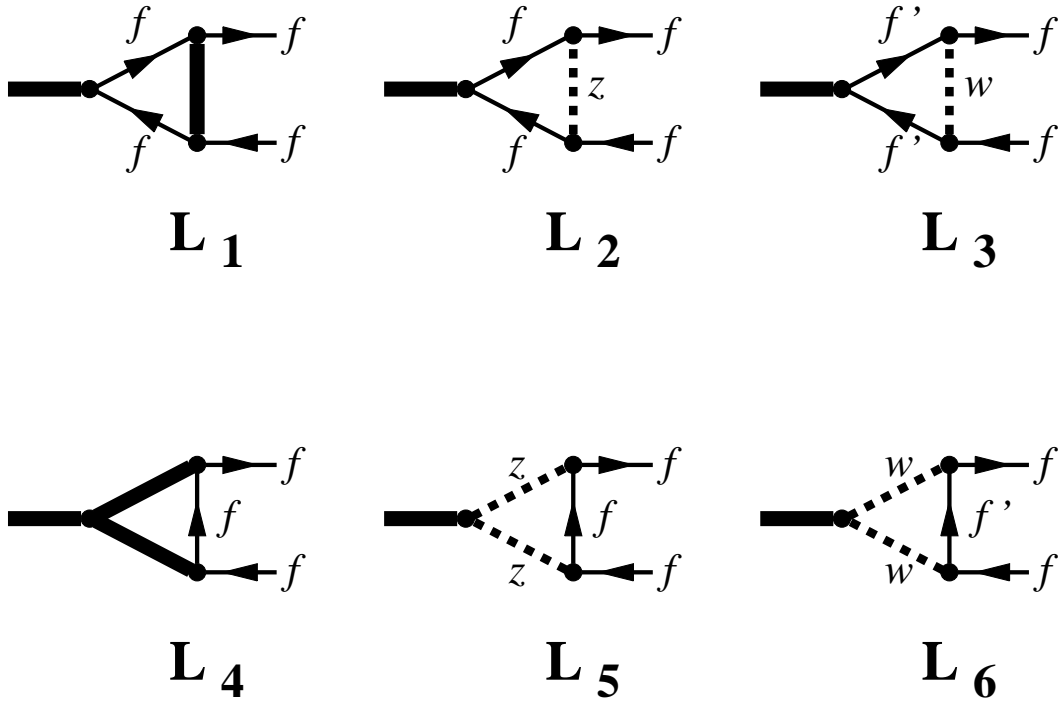}}
}
\vspace{1.5in}
\caption{The six triangle diagrams contributing to $H\rightarrow
  f\bar{f}$ at one loop within the framework of the equivalence theorem.  The
  different lines have the same meaning as in
  Fig.~\protect\ref{figtadpoles}.  The fermion $f'$ is the SU(2)$_L$ partner of
  $f$.}
\label{figtriangles}
\end{figure}

According to the discussion above, the matrix element for the decay
$H\rightarrow f\bar{f}$ is given by the expression
\begin{equation}
-i{\cal M}_{H\rightarrow f\bar{f}}=Z_H^{1/2}
\bar{u}(p_1-p_2,m_f)\left(Z_R^{1/2}P_L+Z_L^{1/2}P_R\right)
\Gamma_3^0\left(Z_R^{1/2}P_R+Z_L^{1/2}P_L\right)v(p_2,m_f).
\label{M}
\end{equation}
Here $p_1$ and $p_2$ are the momenta of the incoming Higgs boson and the
outgoing antifermion, respectively, and $\Gamma_3^0$ is the truncated
three-point vertex function calculated using the bare Lagrangian,
\begin{equation}
\Gamma_3^0=-\frac{i}{\!\sqrt{2}}g_f^0+\sum_{i=1}^6L_i
+\cdots,
\label{gamma3}
\end{equation}
where the one-loop integrals $L_i$ correspond to the triangle
diagrams in Fig.\ \ref{figtriangles}.

To one-loop, the renormalization constants multiply only the leading term in
$\Gamma_3^0$.  Using the definition of $g_f^0$ in Eq.\ (\ref{deltagf}) with
$Z_\Phi=Z_w$ as established above, writing $Z_i$ as $1+\delta Z_i$, and
expanding, we obtain the one-loop expression for
${\cal M}_{H\rightarrow f\bar{f}}$,
\begin{eqnarray}
-i{\cal M}_{H\rightarrow f\bar{f}}
&=&\bar{u}(p_1-p_2,m_f)\left[-\frac{i}{\!\sqrt{2}}g_f^0
\left(1+\frac{1}{2}\delta Z_H-\frac{1}{2}\delta Z_w+\frac{1}{2}\delta Z_L^f
+\frac{1}{2}\delta
Z_R^f+\frac{\delta g_f}{g_f}\right) \right.\nonumber \\
&&\qquad\qquad +\left.\sum_{i=1}^6L_i\;\right]v(p_2,m_f). \label{Mexpanded}
\end{eqnarray}
The ``$\delta$'' contributions in this expression would appear as
renormalization counterterms in a calculation based on physical
rather than bare fields and, for convenience, we will refer to them as such.

It will be useful to split the counterterms into ``bosonic'' and ``fermionic''
parts, and rewrite the decay matrix element to one loop as
\begin{equation}
-i{\cal M}_{H\rightarrow f\bar{f}}
\;=\;\bar{u}(p_1-p_2,m_f)\left[-\frac{i}{\sqrt{2}}g_f\left(
1+L_{bos}+L_{fer}\right)+\sum_{i\;=\;1}^6L_i\right]v(p_2,m_f).
\label{Moneloop}
\end{equation}
where
\begin{equation}
L_{ bos}\;=\;\frac{Z_H^{1/2}}{Z_w^{1/2}}-1\;=\;\frac{1}{2}\delta
Z_H-\frac{1}{2}\delta Z_w+\cdots, \label{Lbos}
\end{equation}
and
\begin{equation}
L_{fer}\;=\;\left(1+\frac{\delta g_f}{g_f}\right)\left(Z_R^fZ_L^f
\right)^{1/2}-1\;=\;\frac{1}{2}\delta Z_R^f+\frac{1}{2}\delta Z_L^f
+\frac{\delta g_f}{g_f}+\cdots. \label{Lfer}
\end{equation}
This grouping of terms has the advantage that $L_{bos}$, while dependent on the
fermions through loop contributions, is independent of the flavor of the final
fermion pair. It therefore gives a universal correction to all decays
$H\rightarrow f\bar f$.  The fermionic counterterm, in contrast, depends
explicitly on the flavor of the final pair.  This distinction will be important
later.

\subsection{The bosonic counterterm}
\label{sub.boson}

\begin{figure}[t]
\vspace*{33pt}
\centerline{
\epsfysize=4.4in {\epsffile{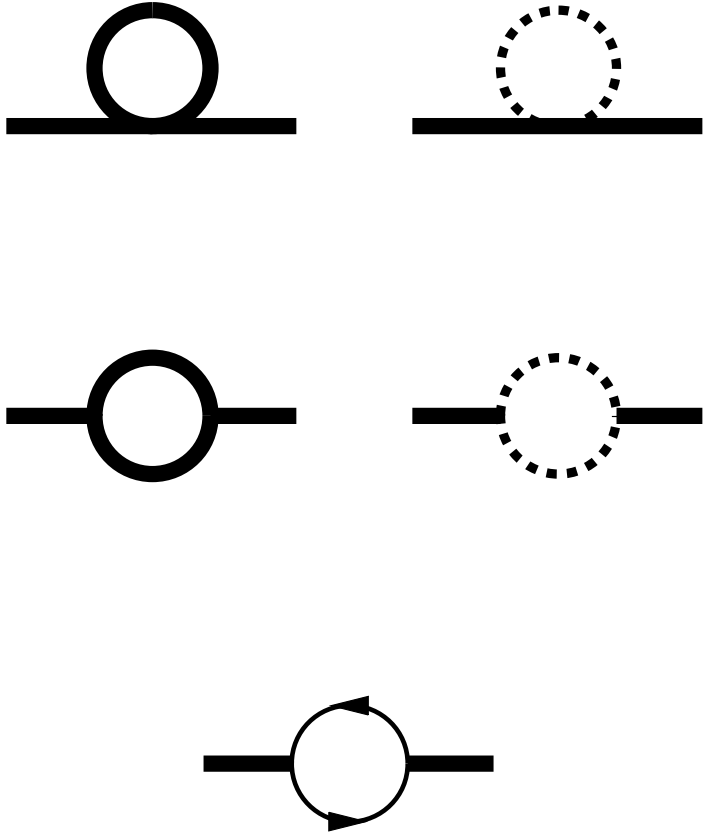}}
}
\vspace{1.0in}
\caption{Higgs self-energy contributions at one loop.
  The different lines have the same meaning as in
  Fig.~\protect\ref{figtadpoles}.
  A summation over all Goldstone boson
  and fermion loops is implied.  }
\label{fighiggsself}
\end{figure}

\begin{figure}[t]
\vspace*{33pt}
\centerline{
\epsfysize=2.8in {\epsffile{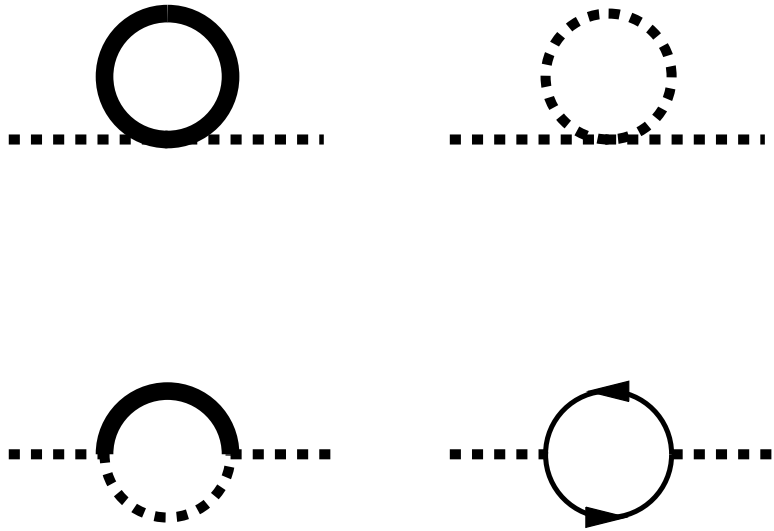}}
}
\vspace{1.1in}
\caption{The Goldstone boson self energies at one loop.
  The different lines have the same meaning as in
  Fig.~\protect\ref{figtadpoles}.  The diagrams on the right need to be summed
  over the different Goldstone boson and fermion loop contributions,
  respectively. In case of an external $w$ or $z$, the fermion loop
  consists of an $ff'$ or $ff$ pair, respectively, where
  $f'$ is the SU(2)$_L$ partner of $f$.  }
\label{figgbself}
\end{figure}

The bosonic counterterm $L_{bos}$ defined in Eq.~(\ref{Lbos}) is determined by
the derivatives of the self-energies $\Pi_{w}^0$ and $\Pi_H^0$ in Eqs.\
(\ref{zw}) and (\ref{zh}).  Because the field renormalization constant $Z_H$
differs from $Z_{w}=Z_\Phi$ only by a finite renormalization, the bosonic
counterterm $L_{bos}$ is finite to all orders.
The tadpole contributions to the self energies are cancelled by the same
counterterm $T_0=v_0\delta m_0^2$ as cancels the apparent radiative changes
in the physical vacuum expectation value $v$, and will be dropped.
The remaining contributions to the boson self energies are given to
one loop by the one-particle irreducible
diagrams shown in Figs.\ \ref{fighiggsself} and \ref{figgbself}. The presence
of fermion loops in these diagrams leads to a breaking of the
SO(3) symmetry of the Higgs Lagrangian for unequal masses of the fermions in an
SU(2)$_L$ doublet, with the result that $\Pi_z^0(p^2)\not=\Pi_w(p^2)$ for
$p^2\not=0$, hence that $Z_z\not=Z_w$.

The boson-loop diagrams have been calculated by a number
of authors \cite{passarino1,marciano,djl1,passarino2,
djl2,dawson,maher,denner}. With the notation
\begin{equation}
\delta Z_i\;=\;\delta Z_{i,bos}+\delta Z_{i,fer}, \label{zbosfer}
\end{equation}
the results are
\begin{eqnarray}
\delta Z_{H,bos}&=&\frac{\lambda}{16\pi^2}\left(12-2\pi
\sqrt{3}\right) ,\label{deltazhbos}\\
\delta Z_{w,bos}&=&\frac{\lambda}{16\pi^2}(-1). \label{deltazwbos}
\end{eqnarray}
These one-loop quantities are separately finite.  At two loops, the $Z$'s
become singular \cite{maher,vanderbij,ghinculov}, but the ratio
$Z_{H,bos}/Z_{w,bos}$ and hence its contribution to $L_{bos}$ remains finite.

The contributions to $\delta Z_H$ and $\delta Z_w$ from the fermionic loops are
\cite{riesselmann}
\begin{eqnarray}
\delta Z_{H,fer}&=&\frac{1}{8\pi^2v^2}\sum_{f}N_C^fm_f^2
        {\rm Re}\!\left[ -\,B_0(M_H^2,m_f^2,m_f^2) \right.\nonumber\\
&&      \qquad\left. - (M_H^2-4m_f^2)\partial B_0(M_H^2,m_f^2,m_f^2)
       \right], \label{deltazhfer}\\
\delta Z_{w,fer} &=&\frac{1}{8\pi^2v^2}\sum_{(f,f')}N_C^f
        {\rm Re}\!\left[-\,(m_f^2+m_{f'}^2)B_0(0,m_{f}^2,m_{f'}^2)
\right.\nonumber\\
&&    \qquad\left. +(m_f^2-m_{f'}^2)^2\, \partial B_0(0,m_f^2,m_{f'}^2)
        \right], \label{deltazwfer}
\end{eqnarray}
where the standard scalar integral $B_0(p^2,m_0^2,m_1^2)$ and its derivative
$\partial B_0(p^2,m_0^2,m_1^2)$ \cite{denner,thooft} are defined in Appendix
\ref{app.int}.
In these expressions, $f'$ is the doublet partner of the fermion $f$.  The
color factor $N_C^f$ is 3 for quarks and 1 for leptons.  While the function
$B_0(p^2,m_0^2,m_1^2)$ and the individual $\delta Z$'s are ultraviolet
divergent, the divergencies cancel in the difference $\delta Z_{H,fer}-\delta
Z_{w,fer}$, and the fermionic contribution to $L_{bos}$ is also finite, as
expected.

Because the overall coupling factors are proportional to $m_f^2/v^2$, light or
massless fermions do not contribute significantly to the radiative corrections.
However, the contributions from top-quark loops can be significant, especially
if $M_H\approx {\rm O}(m_t)$.  We will extract these corrections explicitly.
Neglecting all fermion masses except for $m_t$, the fermionic
contribution to $L_{bos}$ is
\begin{eqnarray}
\frac{1}{2}\delta Z_{H,fer}-\frac{1}{2}\delta Z_{w,fer} \;=\;
 \frac{1}{16\pi^2}N_C^f\frac{m_t^2}{v^2}
\!\!\!\!&\left[  \rule{0pt}{14pt} \right. &\!\!\!\!
-\; {\rm Re}\left[ B_0(M_H^2,m_t^2,m_t^2) - B_0(0,0,m_t^2) \right]
\nonumber \\
\!\!\!\!\!&&\!\!\!\!\left. - \left(1 - \frac{1}{a^2}\right)
        M_H^2\, {\rm Re}\left[ \partial B_0(M_H^2,m_t^2,m_t^2) \right]
 - \frac{1}{2} \;
\right].\quad
\label{lbosfer}
\end{eqnarray}

The explicit expression for $L_{bos}$
depends on the ratio of the Higgs-boson and top-quark masses. For Higgs-boson
masses above
the top-quark production threshold at $M_H=2m_t$, we find that
\begin{eqnarray}
\label{ctbosexactheavy}
L_{bos} \!\!\!\!\!&=& \!\!\!\!\!
\frac{\lambda}{16\pi^2}\left(\frac{13}{2} - \pi\sqrt{3}\right)
                                                                \nonumber\\
           &&\mbox{}+ \frac{3}{16\pi^2}\frac{m_t^2}{v^2}
        \left[\rule{0pt}{18pt}
                b \left( 2 + \frac{1}{a^2} \right)
                       \left( \ln 2a + \ln\left(\frac{1+b}{2}\right)\right)
                - \frac{1}{2} - \frac{1}{a^2}
        \right]\,,\quad M_H>2m_t,
\end{eqnarray}
while for smaller masses, $M_H<2m_t$,
\begin{eqnarray}
\label{ctbosexactlight}
L_{bos} \!\!\!\!\!&=&\!\!\!\!\!
\frac{\lambda}{16\pi^2}\left(\frac{13}{2} - \pi\sqrt{3}\right)
                                                                \nonumber\\
        &&\mbox{}+ \frac{3}{16\pi^2}\frac{m_t^2}{v^2}
          \left[\rule{0pt}{18pt}
                \frac{1}{2a^4}\phi\sin(\phi)
              - \left(2a^2 - 3 + \frac{1}{a^2} \right) \frac{\phi}{\sin(\phi)}
              - \frac{1}{2} - \frac{1}{a^2}
          \right]\,,\; M_H<2m_t,
\end{eqnarray}
where
\begin{equation}
a=M_H/2m_t, \qquad b=\sqrt{1-a^{-2}}, \qquad \phi = \arccos(1-2a^2),\quad
0<\phi<\pi\,.
\end{equation}
We will henceforth refer to the cases $M_H>2m_t$ ($M_H<2m_t$) as those of a
``heavy'' (``light'') Higgs boson.

As noted earlier, the result for $L_{bos}$ does not depend on the final state
of the decay. For a fixed value of $M_H$, it is the same for all processes
$H\rightarrow f\bar f$.  As an useful example, we note that $L_{bos}$ already
gives the {\em complete} one-loop ${\rm O}(\lambda)$ and ${\rm O}(g_t^2)$
corrections to the dominant leptonic decay, $H\rightarrow \tau^+\tau^-$. The
fermionic counterterm and the vertex corrections for this and other leptonic
decays do not receive further corrections in these couplings.

It is worth mentioning that there are no threshold singularities in $L_{bos}$.
The right-hand sides of Eqs.\ (\ref{ctbosexactheavy}) and
(\ref{ctbosexactlight}) are equal for $M_H=2m_t$.  In general, threshold
singularities occur only when the gauge couplings of the Standard Model are
included. We can consistently neglect the gauge couplings when using the
equivalence theorem.

\subsection{Fermionic counterterm}
\label{sub.fermion}

\begin{figure}[t]
\vspace*{33pt}
\centerline{
\epsfysize=1.25in {\epsffile{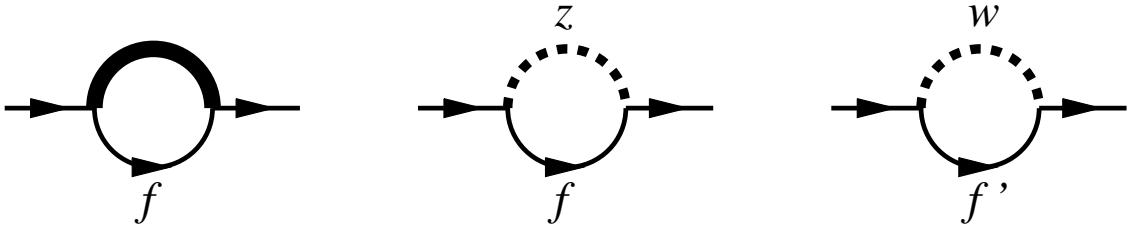}}
}
\vspace{1.3in}
\caption{The three diagrams contributing to the fermion self-energies
  at one loop.  The different lines have the same meaning as in
  Fig.~\protect\ref{figtadpoles}. The fermion $f'$ is the SU(2)$_L$ partner of
  the external fermion~$f$.
  }
\label{figfermiself}
\end{figure}

The fermionic counterterm $L_{fer}$ defined in Eq.\ (\ref{Lfer})
can easily be reduced to the form
\begin{equation}
L_{fer}={\rm Re}\left\{\Pi^0_{S,f}(m_f^2)-2m_f^2\left[
\Pi^{0\,'}_{V,f}(m_f^2)+\Pi^{0\,'}_{S,f}(m_f^2)\right]\right\}
\end{equation}
by using the renormalization conditions stated in Eqs.\ (\ref{zl1}),
(\ref{zr1}), and (\ref{dg1}). The potential tadpole contributions to the
fermionic self energies and their derivatives are exactly cancelled by the
counterterms.
The remaining one-particle irreducible diagrams are shown in Fig.\
\ref{figfermiself}. All involve loops containing either a virtual Higgs boson
or a massless Goldstone boson and the appropriate fermion line, $f$ or $f'$,
where $f$ refers to the final fermion.

In contrast to the expressions for the bosonic self-energies, the expressions
for the fermion self-energies depend on the Dirac matrices and, in particular,
on $\gamma^5$. This $\gamma^5$ dependence raises the possibility of problems
with dimensional regularization related to the definition of such quantities as
${\rm Tr}\gamma^5\gamma^{\mu_1}\cdots\gamma^{\mu_D}$ and the antisymmetric
tensor in $D$ dimensions. However, as shown by Barbieri {\em et al.}
\cite{barbieri}, naive dimensional regularization with $\gamma^5$ and the
remaining Dirac matrices treated as anticommuting is equivalent in the present
context to the proper 't\,Hooft-Veltman scheme \cite{thooft} to at least two
loops for physical quantities. We have therefore calculated the fermion self
energies and the triangle diagrams using naive dimensional regularization.
Using the results in the appendices, we find that
\begin{eqnarray}
\label{ctfer}
\lefteqn{L_{fer} = \frac{1}{16\pi^2}\frac{m_f^2}{v^2}{\rm Re}\!\left\{
\left[
B_0(m_f^2,m_f^2,M_H^2) - B_0(m_f^2,m_f^2,0) - \frac{2m_{f'}^2}{m_f^2}
B_0(m_f^2,m_{f'}^2,0)
\right]\right. }\nonumber\\
&&-2m_f^2\left[
- \partial B_1(m_f^2,m_f^2,M_H^2) - \partial B_1(m_f^2,m_f^2,0) -\left(
1+\frac{m_{f'}^2}{m_f^2}\right)
\partial B_1(m_f^2,m_{f'}^2,0)\right. \nonumber\\
&&\qquad\qquad\left. \left. + \partial B_0(m_f^2,m_f^2,M_H^2)
- \partial B_0(m_f^2,m_f^2,0) -
\frac{2m_{f'}^2}{m_f^2}\partial B_0(m_f^2,m_{f'}^2,0)\right]\right\}.
\end{eqnarray}
The explicit evaluation of this expression requires the specification
of the flavor of the final state fermion. The results for $f=t$ and
$f=b$ are given in Sect.~(\ref{sect.finalhtt}) and
(\ref{htobbar.final}), respectively.

\subsection{The one-loop vertex diagrams}
\label{sub.irred}

The six one-loop vertex diagrams $L_i,\, i$=1--6, are
shown in Fig.\ \ref{figtriangles}. Their contributions are given by
\begin{eqnarray}
L_1 &=& \left(\frac{m_f}{v}\right)^3 \,T_I(p_1^2,p_2^2,m_f^2,m_f^2,M_H^2)
\,,\\
L_2 &=& -\left(\frac{m_f}{v}\right)^3 \gamma^5 \,T_I(p_1^2,p_2^2,m_f^2,m_f^2,0)
\,\gamma^5\,,\\
L_3 &=& \frac{2m_{f'}}{v^3}
        \left(m_f P_L- m_{f'}P_R\right)
 T_I(p_1^2,p_2^2,m_{f'}^2,m_{f'}^2,0)
        \left(P_R\,m_f - P_Lm_{f'}\right) \,,\\
L_4 &=& 6\lambda v\left(\frac{m_f}{v}\right)^2
        \,T_{II}(p_1^2,p_2^2,m_f^2,m_f^2,M_H^2) \,,\\
L_5 &=& -2\lambda v\left(\frac{m_f}{v}\right)^2 \gamma^5
\,T_{II}(p_1^2,p_2^2,0,0,m_f^2) \,\gamma^5\,,\\
L_6 &=& 2\lambda v\frac{1}{v^2} \left(m_fP_L - m_{f'}P_R\right)
 T_{II}(p_1^2,p_2^2,0,0,m_{f'}^2) \left(P_Rm_f - P_Lm_{f'}\right)\,.
\end{eqnarray}
The functions $T_I$ and $T_{II}$ are defined by the
the integrals
\begin{eqnarray}
T_I(p_1^2,p_2^2,m^2,m^2,M^2) &=&\nonumber\\
(2\pi\mu)^{4-D}\int\frac{d^Dq}{(2\pi)^4}\!\!\!\!\!&\!\!&\!\!\!\!\!\!\!
\frac{\not\!p_1\not\!q \,+ m\!\not\!p_1 \,+ \not\!q\not\!q
\,+ 2\,m\!\!\not\!q \,+ m^2}
{ [q^2-m^2+i\epsilon'][(q+p_1)^2-m^2+i\epsilon'][(q+p_2)^2-M^2+i\epsilon'] },
\label{t_i}\\
T_{II}(p_1^2,p_2^2,m^2,m^2,M^2) &=&\nonumber\\
(2\pi\mu)^{4-D}\int\frac{d^Dq}{(2\pi)^4}\!\!\!\!\!&\!\!&\!\!\!\!\!\!\!
\frac{-\not\!q \,\,+ \not\!p_2 + M }
        { [q^2-m^2+i\epsilon'][(q+p_1)^2-m^2+i\epsilon'][(q+p_2)^2-M^2
+i\epsilon'] }.\label{t_ii}
\end{eqnarray}
These tensor integrals can be evaluated using standard techniques
\cite{denner,pasvelt}.

While the triangle functions $L_i,\, i\!=\!2,3,5,6$ depend individually on
$\gamma^5$, the $\gamma^5$'s drop out in the sum of these four one-loop
contributions. The decay matrix element ${\cal M}_{H\rightarrow f\bar{f}}$\,,
Eq.~(\ref{Mexpanded}), has no $\gamma^5$ terms. The remaining Dirac matrices
can be eliminated from the expression for ${\cal M}_{H\rightarrow f\bar{f}}$ by
using the Dirac equations for the spinors $\bar{u}(p_1-p_2,m_f)$ and
$v(p_2,m_f)$. Extracting a factor $-im_f/v$ from the reduced expressions for
the functions $L_i$ and denoting the results by $\widetilde{L_i}$, we can then
write Eq.~(\ref{Moneloop}) as
\begin{equation}
-i{\cal M}_{H\rightarrow f \bar f} = -i\frac{m_f}{v}
\bar u(p_1-p_2,m_f)\, v(p_2,m_f)\,
(1+\Delta {\cal T})\, ,
\end{equation}
where the spinor matrix element is purely scalar and the
quantity $\Delta {\cal T}$ is defined as
\begin{equation}
\label{finalexact}
\Delta {\cal T}
  =  L_{bos} + L_{fer} + L_{tri},\qquad
L_{tri}=\sum_{i=1}^{6}\widetilde{L_i}\, .
\end{equation}
The correction to the decay width can then be written as
\begin{eqnarray}
\label{corrhff}
\Gamma\left(H\rightarrow f\bar f\,\right)
&=&\Gamma_{\rm B}\left({H\rightarrow f\bar f\,}\right)
\left|\,1+\Delta{\cal T}\,\right|^2 \nonumber\\
&=&\Gamma_{\rm B}\left({H\rightarrow f\bar f\,}\right)
\left[\,1 + 2\, {\rm Re}\Delta
{\cal T} +O\left(\lambda^2,\lambda g_f^2,g_f^4\right)\,\right]\,,
\end{eqnarray}
\noindent where \cite{born}
\begin{equation}
\Gamma _B\left(H\rightarrow f\bar f\,\right)=
{N_cm_f^2M_H\over8\pi v^2}\left(1-{{4{m_f^2}}\over {M_H^2}}\right)^{3/2}
\label{eqborn}
\end{equation}
is the Born result.  The function $\sum_i\widetilde{L}_i$ is rather
complicated, and we will only give the results needed for the decays
$H\rightarrow t\bar{t}$ and $H\rightarrow b\bar{b}$. They are listed in the
following sections.

\subsection{$H\rightarrow t\bar{t}$}
\label{sect.finalhtt}

In the mass region $M_H>2m_t$, the only important fermionic decay of the Higgs
particle is the decay into a pair of top quarks, with a branching ratio of
approximately 10\% \cite{branch}.  Since $m_t\gg m_b$, we can calculate the
decay rate for $H\rightarrow t\bar{t}$ using our general results for
$H\rightarrow f\bar f$ evaluated in the limit $m_b$=0.  To simplify the
expressions, we again use the notation
\begin{equation}
a=\frac{M_H}{2m_t}\,\qquad b = \sqrt{1-\frac{1}{a^2}}\, .
\end{equation}
Using the relation $\lambda = M_H^2/(2v^2)$, we add the contributions from the
six triangular loops and obtain
\begin{eqnarray}
\label{allloops}
L_{tri} &=& -\frac{1}{16\pi^2}\left(\frac{m_t}{v}\right)^2
\frac{4a^4}{a^2-1}
 \left[\: \left( 1 + \frac{2}{a^2} - \frac{1}{a^4}\right)\,
                        m_t^2\,C_0(M_H,m_t,m_t,m_t,M_H)  \right. \nonumber \\
&& +\left( 9 - \frac{6}{a^2} \right)\,m_t^2\,C_0(M_H,m_t,M_H,M_H,m_t)
- m_t^2 \,C_0(M_H,m_t,0,0,m_t)\nonumber \\
&&-\frac{1}{2a^2}\,m_t^2\,C_0(M_H,m_t,0,0,0)
+\left(-\, 6 +\frac{7}{a^2} \right)\,\ln 2a\nonumber \\
&& -b\left(-\,6+\frac{2}{a^2}+\frac{1}{a^4}\right)\left(\ln 2a+\ln\frac{1+b}{2}
\right)
+\left. \frac{\pi\sqrt{3}}{2a^2}
-\frac{i\pi}{2a^4}\left(a^2-b\right)\: \right] \, .
\end{eqnarray}
This result, based purely on the Lagrangian of the Higgs sector and the
interaction of the Higgs and the Goldstone bosons with the fermion sector, is
in complete agreement with the corresponding result\footnote{The result given
  in Eq.~(\ref{allloops}) should be compared with the quantity
  $\frac{\alpha}{4\pi}\delta_{weak}$ as defined by Kniehl.} given by Kniehl
\cite{kniehl}, which was calculated using the {\it full} electroweak
Lagrangian, assuming $M_H^2 > 4m_t^2 \gg M_W^2 \gg m_b^2$ and neglecting terms
that are not enhanced by an inverse power of $M_W^2$. The latter is equivalent
to the limit $g_2\rightarrow 0$, with $g_2 /M_W=2/v$.  The agreement of the two
results demonstrates the validity of the Goldstone boson equivalence theorem to
one loop for $H\rightarrow t\bar t$.

To compare the decay matrix element of our equivalence-theorem calculation and
the full electroweak calculation, we still need explicit expressions for
the counterterms $L_{bos}$ and $L_{fer}$ for $M_H>2m_t$. The expression for
$L_{bos}$ is given in Eq.~(\ref{ctbosexactheavy}).
Evaluating the general expression for $L_{fer}$ in
 Eq.~(\ref{ctfer}) for $M_H>2m_t$ using $m_b=0$, we find that
\begin{eqnarray}
\label{ctferexact}
   L_{fer} &=& \frac{1}{16\pi^2}\frac{m_t^2}{v^2}\Bigl[\,
        \left(32a^4 - 36 a^2 + 4\right)\,\ln 2a - 8a^2 + 7
   \nonumber\\
&&\phantom{  \frac{1}{16\pi^2}\frac{m_t^2}{v^2}}
        - \left. b\left(32a^4 - 20a^2\right) \left(\ln 2a
+\ln\frac{1+b}{2}\right)\,\right]\, .
\end{eqnarray}
$L_{fer}$ includes terms proportional to $a^4 \ln 2a$, $a^4$, $a^2 \ln 2a$, and
$a^2$ which involve powers of $M_H^2/m_t^2$ and are enhanced for large Higgs
masses. However, these apparent enhancement terms cancel exactly when we expand
the quantity $b$ in terms of $1/a^2$, that is, $L_{fer}$ is actually {\it not}
enhanced in powers of $M_H$.  This is a manifestation of the Veltman screening
theorem \cite{mveltman}: at one loop, an internal Higgs-boson line with a large
mass $M_H$ leads only to a logarithmic enhancement. Since $L_{fer}$ is derived
from two-point functions with external fermion lines, no power-like enhancement
is possible.

Assembling the partial results according to Eq.~(\ref{finalexact}), we find the
complete one-loop result for $\Delta{\cal T}$ in the limit $m_t\gg m_f$ for
$f\not=t$,
\begin{eqnarray}
\label{httexact}
\Delta {\cal T}
        &=& -\frac{1}{16\pi^2}\left(\frac{m_t}{v}\right)^2
        \left\{ \frac{a^2}{a^2-1}
        \left[ \left( 1 + \frac{2}{a^2} - \frac{1}{a^4}\right)\,
        M_H^2\,C_0(M_H,m_t,m_t,m_t,M_H)  \right.\right. \nonumber \\
&& +\left( 9 - \frac{6}{a^2} \right)\,M_H^2\,C_0(M_H,m_t,M_H,M_H,m_t)
-M_H^2 \,C_0(M_H,m_t,0,0,m_t)\nonumber \\
&&-\frac{1}{2a^2} M_H^2\,C_0(M_H,m_t,0,0,0)
+\left. 2\pi\sqrt{3}
-\frac{2i\pi}{a^2}\left(a^2-b\right)\right]\nonumber\\
&&
+ 8a^2  -\frac{11}{2} +\frac{3}{a^2}
+\left(-32a^4 + 12a^2  +\frac{4}{a^2-1} \right)\,\ln (2a)\nonumber \\
&& + b\left(32a^4 +4a^2+\,10-\frac{3}{a^2}+\frac{12}{a^2-1}
\right)\left(\ln (2a)+\ln\frac{1+b}{2}\right) \nonumber \\
&&\left.
-2a^2\left(\frac{13}{2} - \pi\sqrt{3}\right)\:
\right\},\label{deltaTttbar}
\end{eqnarray}
where we have replaced an overall factor $4a^2m_t^2$ by $M_H^2$ in the
coefficients of the $C_0$ functions.  In the context of the equivalence theorem
this result is exact except for the approximation $m_f=0$, $f\not=t$. The
correction to the decay width is
\begin{eqnarray}
\Gamma\left(H\rightarrow t\bar t\right)
&=&\Gamma_{\rm B}\left({H\rightarrow t\bar t}\right)
\left[1 + 2\, {\rm Re}\Delta
{\cal T} +O\left(\lambda^2,\lambda g_t^2,g_t^4\right)\right].
\end{eqnarray}
%
%

The functions $M_H^2\,C_0$ in Eq.\ (\ref{deltaTttbar}) are functions only of
the ratio $M_H^2/m_t^2$, i.e., of $a^2$.  Expanding these functions for $a^2\gg
1$, we find that all of the contributions to $\Delta{\cal T}$ which are
proportional to positive powers of $a^2$ cancel except for the last term in
Eq.\ (\ref{deltaTttbar}). That term arises from the boson-loop contribution to
$L_{bos}$.  The contributions from the fermion renormalization constants and
the three-point functions are not power-enhanced, but grow only as $\ln
(2a)=\ln (M_H/m_t)$.  The purely bosonic corrections therefore give the
dominant contribution to $\Delta{\cal T}$ for large Higgs-boson mass, a result
that remains true to all orders in perturbation theory.  In particular, at one
loop,
\begin{equation}
\Delta{\cal T}\sim \frac{M_H^2}{16\pi^2v^2}\left[\,\frac{1}{4}
\left( 13-2\pi\sqrt{3}\right)+
{\rm O}\left(\,\frac{m_t^2}{M_H^2}\ln\frac{M_H^2}{m_t^2}\,\right)
+{\rm O}(\,\frac{m_t^2}{M_H^2}\,)\,\right]\,,
\qquad M_H^2\gg m_t^2,\label{asymp}
\end{equation}
and the {\em fraction} of the total correction associated with the top-quark
Yukawa coupling decreases rapidly for $M_H\gg m_t$. However, the actual
correction associated with $g_t$ may still be significant.

\begin{figure}[t]
\vspace*{13pt}
\centerline{
\epsfysize=4.8in \rotate[l]{\epsffile{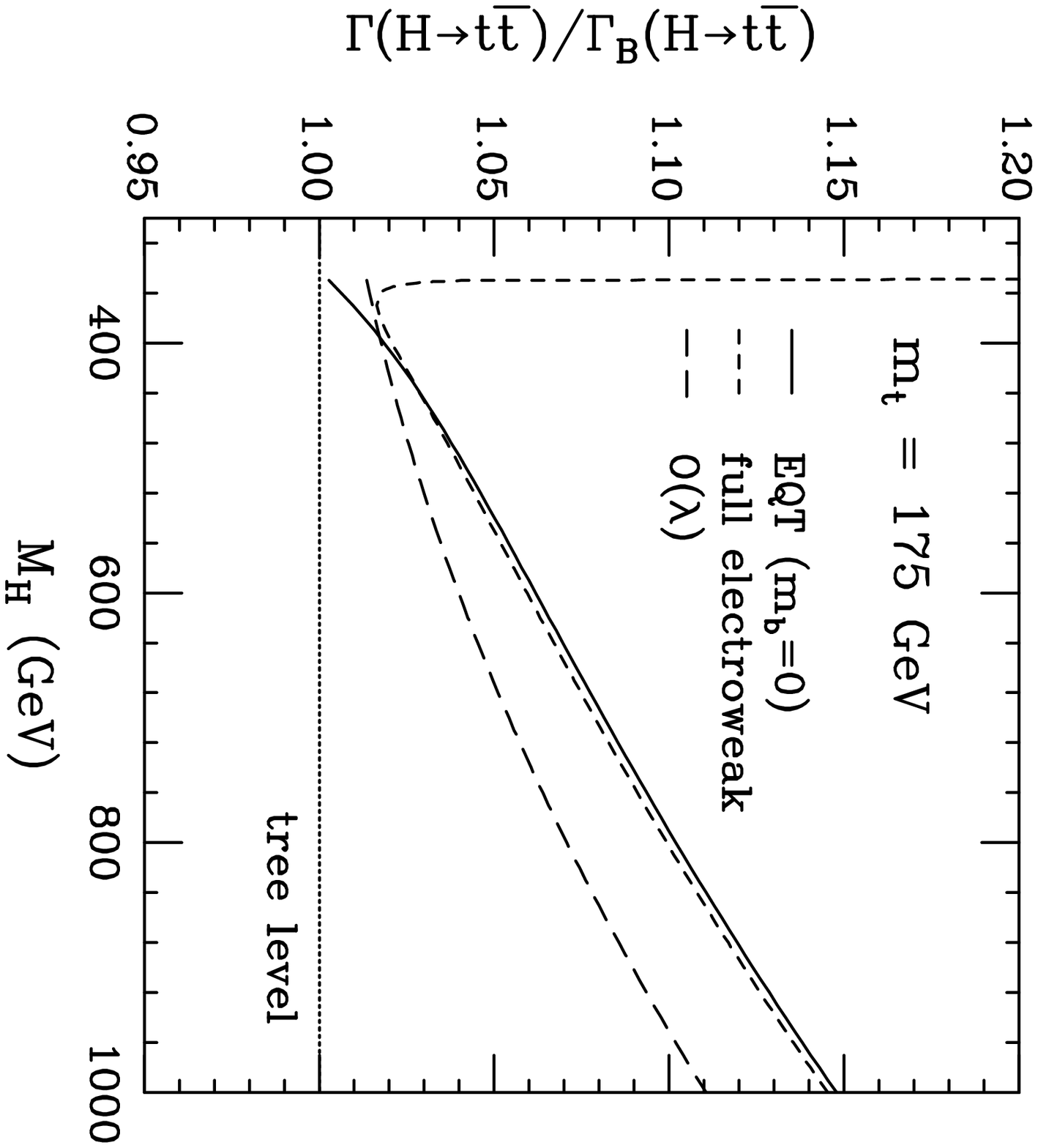}}
}
\vspace{0.15in}
\caption{The one-loop correction factor to the decay width
  $H\rightarrow t\bar t$. The solid curve gives the
  equivalence theorem (EQT) result consisting of the sum of the
  ${\rm O}(\lambda)$ and ${\rm O}(g_t^2)$ corrections. Light
  fermion couplings are neglected. The result is compared with the full
  electroweak correction obtained in \protect\cite{kniehl,hollik,bardin} and
  the EQT result without fermion corrections.
  }
\label{fightteqt}
\end{figure}

In Fig.~\ref{fightteqt} we show the equivalence theorem correction to the decay
width, $\Gamma/\Gamma_B = 1+2\,{\rm Re}\Delta{\cal T}$,
for $H\rightarrow t \bar{t}$ in the limit $m_b=0$ (solid curve). This
result is compared with the full electroweak correction (short dashes)
\cite{hfflong}. Away from the threshold, the equivalence theorem provides
an excellent approximation. For $M_H=500$ GeV (1~TeV), the EQT result is only
3.9\% (1.8\%) larger in magnitude than the full electroweak result.
This is roughly the accuracy
one would expect for a gauge contribution to $\Delta\cal T$ of order
$\alpha/\pi$.


We find that the ${\rm O}(\lambda)$ correction (long dashes) represents the
dominant contribution to the EQT correction for all values of $M_H$ larger than
$2m_t$, i.e., above the decay threshold.  The one-loop Yukawa correction, ${\rm
  O}(g_t^2)$, vanishes at about $M_H=400$ GeV.  For larger values of $M_H$, it
is positive and adds to the ${\rm O}(\lambda)$ contribution.  While it
initially grows more rapidly as a function of $M_H$ than the ${\rm O}(\lambda)$
correction, the $M_H^2$ behavior of the latter wins over the asymptotically
logarithmic growth of the Yukawa correction for $M_H$ greater than about 600
GeV.  The Yukawa contribution represents a decreasing fraction of the total
correction for higher masses, but is still significant numerically.

The very different behaviour of the EQT result and the full electroweak result
at threshold is caused by a Coulomb singularity associated with
the exchange of a virtual photon. Except for this QED
effect, the EQT correction and the weak correction are in qualitative
agreement at $M_H=2m_t$: the EQT correction is about zero at threshold,
and the weak correction is of the order of one percent.  To have a better test
of the validity of the equivalence theorem for values of $M_H\approx 2m_t$, we
next consider the decay $H\rightarrow b\bar{b}$ which is free of the
Coulomb singularity at $M_H=2m_t$.

\subsection{$H\rightarrow b\bar{b}$}
\label{htobbar.final}

In the mass region of a ``light'' Higgs boson, $M_H<2m_t$, the dominant
fermionic decay of the Higgs boson is the (much suppressed) decay $H\rightarrow
b\bar{b}$. We can still treat this decay using the equivalence theorem provided
that $M_H$ is large compared to the masses of the $W$ and $Z$ bosons, and can
use the comparison of the results with those of the full electroweak theory to
test the limits of validity of the equivalence theorem for a ``light'' Higgs
boson.  For a ``heavy'' Higgs boson, $M_H>2m_t$, the equivalence theorem result
and the full electroweak result would be expected to agree to
an accuracy similar to that in the previous section.
However, the dependence of the one-loop correction to
$H\rightarrow b\bar{b}$ on $m_t$ is different from that encountered in
top-quark production and therefore  actually provides an
independent check of the quality of
the equivalence theorem for the case of a heavy Higgs boson.

We will again write the decay matrix element corrected to one loop in
the form
\begin{equation}
-i{\cal M}_{H\rightarrow b\bar{b}}= -i\frac{m_b}{v}\bar{u}(p_1-p_2,m_b)
v(p_2,m_b)\left(1+\Delta{\cal T}\right),\qquad
\Delta{\cal T}=L_{bos}+L_{fer}+L_{tri}. \label{Mbbar}
\end{equation}
The Born result is proportional to the ratio $m_b/v$.  The one-loop
equivalence-theorem corrections contributing to $\Delta{\cal T}$ are
proportional to $\lambda$, $m_t^2/v^2$, $m_b^2/v^2$, and squares of lighter
fermion masses.  Evaluating these contributions to $\Delta{\cal T}$, we set
$m_b=0$ and also neglect other light fermions.  In this limit, $L_{fer}$ only
receives contributions from the bottom-quark self-energy diagram which contains
a $(w,t)$ loop, Fig.\ \ref{figfermiself}, and $L_{tri}$ reduces to a sum over
only two triangle graphs, $L_3$ and $L_6$ in Fig.\ \ref{figtriangles}, with the
tree-level factor $-im_b/v$ extracted.

The bosonic counterterm $L_{bos}$ is given by Eq.\ (\ref{ctbosexactheavy}) or
(\ref{ctbosexactlight}), depending on the value of $M_H$.  The fermionic
counterterm, $L_{fer}$, and the triangle graphs, $L_{tri}$, depend on the
flavor of the final fermion pair and need to be re-evaluated for the present
decay into bottom-quarks, i.e., their results differ from the above results for
the decay of the Higgs boson into top quarks.

The fermionic counterterm is evaluated using the general result in
Eq.\ (\ref{ctfer}) with $m_b\rightarrow 0$,
\begin{equation}
L_{fer}=\frac{1}{16\pi^2}\frac{m_t^2}{v^2}\left[\, -2B_0(0,0,m_t^2)\,\right]\,.
\label{Lfer_b}
\end{equation}

The triangle contribution is calculated from the expressions for $L_3$
and $L_6$ using the Dirac equation and setting $m_b=0$:
\begin{eqnarray}
L_{tri}\!&\!=\!&\!\frac{1}{16\pi^2}\frac{m_t^2}{v^2}
\left[
   2B_0(0,0,m_t^2)
   +\left(2 - \frac{1}{a^2} \right)
            \left[B_0(M_H^2,m_t^2,m_t^2) - B_0(0,0,m_t^2) \right]
   + 2 - 4\ln 2a
\right.\nonumber\\
&&\!\!\!\left.
   +\frac{1}{4a^4}M_H^2\,C_0(M_H^2,0,m_t^2,m_t^2,0)
   +2\left(1+\frac{1}{4a^2}\right)M_H^2\,C_0(M_H^2,0,0,0,m_t^2)\,+ 2i\pi\,
\right]. \label{Ltri_b}
\end{eqnarray}

The divergencies related to the integrals $B_0$ cancel in the sum
$\Delta{\cal T}=L_{bos}+L_{fer}+L_{tri}$. In case of a ``light'' Higgs, the
complete one-loop electroweak radiative correction to the amplitude
for the decay $H\rightarrow b\bar{b}$ is now found to be
\begin{eqnarray}
\Delta{\cal T}&=&\frac{1}{16\pi^2}\frac{m_t^2}{v^2}
\left[
\frac{5}{2} -\frac{4}{a^2}-4\ln 2a
-\left(\frac{1}{a^2}-\frac{2}{a^4}\right)\phi\sin(\phi)
- 3\left( 2a^2 -3 +\frac{1}{a^2}\right) \frac{\phi}{\sin(\phi)}
\right.\nonumber\\
&&
  + \frac{1}{4a^4}M_H^2\,C_0(M_H^2,0,m_t,m_t,0)
  + 2\left(1+\frac{1}{4a^2}\right)M_H^2\,C_0(M_H^2,0,0,0,m_t^2)\nonumber\\
&&\left.
  +2a^2\left(\frac{13}{2}-\pi\sqrt{3}\right) + 2i\pi
\right]\,,\quad M_H<2m_t,
\end{eqnarray}
and the heavy-Higgs result is
\begin{eqnarray}
\Delta{\cal T}&=&\frac{1}{16\pi^2}\frac{m_t^2}{v^2}
\left[
\frac{5}{2} -\frac{4}{a^2}-4\ln 2a
+ b\left(2+\frac{5}{a^2}\right)
       \left[\ln(2a)+\ln\left(\frac{1+b}{2}\right)\right]
\right.\nonumber\\
&&
  + \frac{1}{4a^4}M_H^2\,C_0(M_H^2,0,m_t,m_t,0)
  + 2\left(1+\frac{1}{4a^2}\right)M_H^2\,C_0(M_H^2,0,0,0,m_t^2)\nonumber\\
&&\left.
  +2a^2\left(\frac{13}{2}-\pi\sqrt{3}\right)
+ i\pi\,\left( 2 -b\left(1+\frac{1}{a^2}\right)\right)
\right]\,,\quad M_H>2m_t.
\end{eqnarray}

\begin{figure}[t]
\vspace*{13pt}
\centerline{
\epsfysize=4.8in \rotate[l]{\epsffile{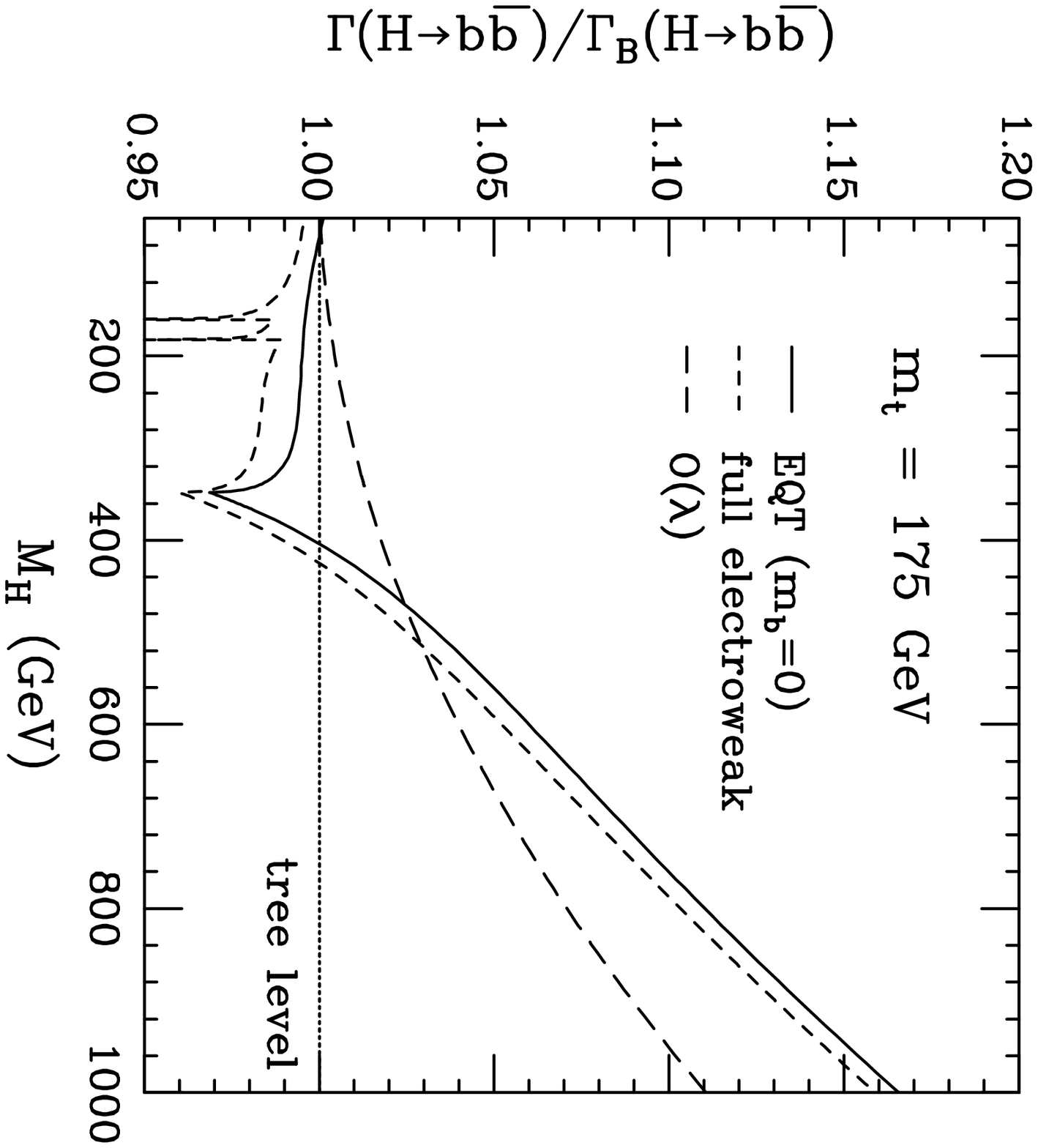}}
}
\vspace{0.15in}
\caption{The one-loop correction factor to the decay width
  $H\rightarrow b\bar b$. The solid curve gives to the equivalence theorem
  result consisting of the sum of the ${\rm O}(\lambda)$ and the
 ${\rm O}(g_t^2)$ corrections. Light
  fermion couplings are neglected. The result is compared with the full
  electroweak correction obtained in \protect\cite{kniehl,hollik,bardin} and
  the EQT result without fermion corrections.
  }
\label{fighbbeqt}
\end{figure}

In Fig.~\ref{fighbbeqt} we show the equivalence theorem correction to the decay
width, $\Gamma/\Gamma_B = 1+2\,{\rm Re}\Delta{\cal T}$, for $H\rightarrow b
\bar{b}$ in the limit $m_b=0$ (solid curve). This result is compared with the
full electroweak correction (short dashes) \cite{hfflong}. For large values of
$M_H$, the EQT result approximates the full correction very well,
in agreement with our findings in the case of $H\rightarrow t\bar{t}$.
The difference between
the two corrections is again of the magnitude of the expected
gauge corrections.  The dominant contribution overall is
the ${\rm O}(\lambda)$ correction (long dashes), which grows
more rapidly than the Yukawa correction for $M_H$ larger than about 700 GeV.

For a Higgs mass of about 400 GeV, the top-quark Yukawa correction cancels the
${\rm O}(\lambda)$ contribution, and the one-loop radiative corrections are
actually determined mainly by the gauge
corrections.  However, the magnitude of the gauge correction is still small
compared to the magnitude of the Yukawa
correction at this point, with an absolute value less than 1\%.
The equivalence theorem breaks down for $M_H$ less than about 200 GeV, where
the condition $M_H\gg M_W,\,M_Z$ is of questionable validity.
The gauge corrections are the dominant
corrections in this region, and are larger in
magnitude than both the Yukawa and the ${\rm O}(\lambda)$
corrections. In addition, the gauge correction displays Coulomb
singularities at $M_H=2M_Z$ and $2M_W$, a feature which is unique to gauge
interactions and cannot be reproduced by the equivalence theorem. This
threshold region requires special treatment.

Qualitatively, the EQT result approximates the full electroweak corrections
rather well for Higgs masses larger than 200 GeV.  The full
electroweak correction remains finite at the threshold for top-quark
production at $M_H=2m_t$, and the expected threshold kink is reproduced by the
equivalence-theorem calculation.
We conclude that the equivalence theorem is a useful tool even in the case of a
``light'' Higgs boson with a mass in the range $2M_Z<M_H<2m_t$,
with the EQT correction to $\Gamma(H\rightarrow b\bar{b})$ giving a good
estimate of the full electroweak correction.

\section{Summary}

The equivalence theorem is known to be an excellent tool in describing heavy
Higgs physics. Calculations based on the equivalence theorem are usually
carried out by using massless Goldstone bosons and a massive Higgs, neglecting
all gauge and Yukawa couplings. While the original applications of the
equivalence theorem were to tree-level processes \cite{lee,chan}, it has
since been shown that the equivalence theorem also provides a simple way of
calculating the dominant contributions of internal
$W^\pm$, $Z$, and $H$ bosons \cite{bagger,hvelt,he}.

In this paper, we have extended the equivalence theorem by systematically
including the Yukawa interactions.  This is possible
because the approximations underlying the
equivalence theorem are independent of the values of the Yukawa couplings.
In particular, we have formulated a renormalization procedure which is
simultaneously consistent
with the requirements of the equivalence theorem, and has the
correct relations to physical observables in the limit of vanishing
gauge couplings. Because the top quark is quite massive, $g_t$ is large,
and it is generally necessary to include the top-quark Yukawa coupling
in calculations of electroweak radiative corrections. The framework
presented here allows these calculations to be done rather simply using
the EQT, with the gauge couplings set to zero. One can hope to
obtain good approximate results for radiative corrections to
Higgs-sector processes in the
case of a ``light'' Higgs, $2M_Z<M_H<2m_t$, and excellent approximations for
larger Higgs masses.

As an example, we calculated the one-loop corrections to fermionic Higgs decays
using the equivalence theorem with fermions, and compared our results with
the results obtained from a full electroweak calculation. Since the Yukawa
interactions are negligible except for the top-quark coupling, we only included
contributions from the latter. We find that the Higgs coupling $\lambda$
and the Yukawa coupling $g_t$ give the dominant corrections to the decay rates
for $M_H>2m_t$. The much smaller contributions of the transverse gauge
couplings are only significant very close to 400 GeV in the case
$H\rightarrow b\bar{b}$ where the
dominant contributions cancel, and near the decay threshold
for $H\rightarrow t\bar{t}$ where virtual photon exchange
produces a Coulomb singularity.

In the mass range $2M_Z<M_H<2m_t$, the process $H\rightarrow b\bar b$ is the
only significant fermionic decay. The one-loop radiative corrections to
this decay associated with the quartic Higgs-boson coupling, the top-quark
Yukawa coupling, and the transverse electroweak gauge couplings are
all similar in magnitude, but with differing signs.  In particular,
a partial cancellation of the Higgs and Yukawa contributions
makes the gauge correction equally important. The total correction
is very small, less than 2\%.
It seems plausible that the sum of the {\it
magnitudes} of Higgs and Yukawa contribution would give a good estimate for an
upper bound on the magnitude of the complete electroweak radiative correction
for a variety of electroweak processes.  In absence of cancellations between
the $O(g_t^2)$ and $O(\lambda)$ corrections, the equivalence theorem result is
expected to be the dominant correction for Higgs masses larger than $2M_Z$.
Threshold singularities arising from the gauge sector cannot, of course, be
reproduced using the equivalence theorem, and require special treatment
in any case.

According to Veltman's screening theorem \cite{mveltman}, the ${\rm
  O}(\lambda)$ corrections are the only one-loop corrections that grow
proportional to $M_H^2$. We find that this asymptotic growth of the correction
is dominant only for Higgs-boson masses larger than 600 to 700 GeV, assuming a
top-quark mass of 175 GeV. For smaller values of $M_H$, we find the ${\rm
  O}(g_t^2)$ corrections to have the stronger dependence on $M_H$.

In conclusion we find that the calculation of radiative corrections using the
equivalence theorem is greatly improved if the Yukawa interactions are
included. The limit of $m_f=0, f\not = t$ allows for a relatively simple
calculation of the dominant radiative corrections, yielding an excellent
approximation of the full electroweak corrections for the heavy Higgs-case,
and order-of-magnitude estimates for $2M_Z < M_H < 2m_t$.

\section*{Acknowledgments}

The authors would like to thank B.~Kniehl for providing the data for the full
electroweak corrections shown in Figs.\ \ref{fightteqt} and
\ref{fighbbeqt}. One of the
authors (K.R.) would like to thank A.~Dabelstein for useful conversations, and
the Deutsche Forschungsgemeinschaft (DFG) for financial support under contract
no.\  Li519/2-1. The work of the other author (L.D.) was supported in part
by the U.S. Department of Energy under Contract No. DE-FG02-95ER40896.


\appendix
\section{Tadpoles and self-energies of $H, z, w$ and $f$.}
\label{app.self}

The only neutral field that receives a shift in its vacuum expectation value
because of tadpole contributions is the Higgs field $h$; the $z$ field does
not. The $z$ and $w$ fields have couplings through ${\cal L}_H$, Eq.\
(\ref{lhiggs}), that require an even number of fields to participate, and no
purely bosonic tadpole graphs can be formed for the $z$.  With the addition of
the fermionic Lagrangian ${\cal L}_F$, Eq.\ (\ref{lfermi}), the $z$ can form a
tadpole with a fermion loop, but the presence of a factor $\gamma^5$ in the
$z$-fermion coupling and a trace over the $\gamma$-matrices involved yields a
vanishing result. However, fermion loops contribute to the Higgs tadpoles as
shown in  Fig.\ \ref{figtadpoles} since no $\gamma^5$ is involved in the
$H$-fermion coupling.  Taking into account both the bosonic and fermionic
contributions, the Higgs one-point function (tadpole function) is
\begin{equation}
\label{tad.eq}
T= \frac{1}{16\pi^2}\left(-\frac{3M_H^2}{2v}A_0(M_H^2)+\sum_{f}N_C^f
\frac{4m_f^2}{v}A_0(m_f^2) \right) \, . \label{tadint}
\end{equation}

The graphs of the one-loop one-particle irreducible self-energy
contributions to the bosonic and fermionic fields are shown
in Figs.\ \ref{fighiggsself}, \ref{figgbself}, and \ref{figfermiself}.
They can be evaluated as \cite{denner,riesselmann}
\begin{eqnarray}
\label{self.eq}
\Pi_H^0(p^2)&=&
-\frac{1}{16\pi^2}\left(3\lambda A_0(M_H^2)+18\lambda^2
v^2B_0(p^2,M_H^2,M_H^2) - 6\lambda^2 v^2B_0(p^2,0,0)\right)\nonumber\\
&&+\frac{1}{16\pi^2}\sum_{f}N_C^f\frac{m_f^2}{v^2}
        \left( 4 A_0(m_f^2) - (2p^2-8m_f^2)B_0(p^2,m_f^2,m_f^2)\right)\,,\\
\Pi_z^0(p^2)&=&
-\frac{1}{16\pi^2}\left(\lambda A_0(M_H^2)+4\lambda^2
v^2B_0(p^2,M_H^2,0) \right)\nonumber\\
&&+\frac{1}{16\pi^2}\sum_{f}N_C^f\frac{m_f^2}{v^2}
        \left( 4 A_0(m_f^2) - 2p^2B_0(p^2,m_f^2,m_f^2) \right)\,,
\label{pizint}\\
\Pi_w^0(p^2)&=&
-\frac{1}{16\pi^2}\left(\lambda A_0(M_H^2)+4\lambda^2
v^2B_0(p^2,M_H^2,0) \right)\nonumber\\
&& -\frac{1}{16\pi^2}\sum_{(f,f')}N_C^f\frac{1}{v^2}\left[
                8m_f^2m_{f'}^2B_0(p^2,m_f^2,m_{f'}^2)  \right.\nonumber\\
&&\phantom{xxxxxxxxxxxx} +2(m_f^2+m_{f'}^2)\left(-A_0(m_f^2)-A_0(m_{f'}^2)
\right.\nonumber\\
&&\phantom{xxxxxxxxxxxxxx}\left. \left.
        +(p^2-m_f^2-m_{f'}^2)B_0(p^2,m_f^2,m_{f'}^2)\right)\right]\,,\\
\Pi_{V,f}^0(p^2) \!\!\!\!\!&=&\!\!\!\!\!
\frac{1}{16\pi^2}\frac{m_f^2}{v^2}\left[
-B_1(p^2,m_f^2,M_H^2) -B_1(p^2,m_f^2,0)-\left(
1+\frac{m_{f'}^2}{m_f^2}\right)B_1(p^2,m_{f'}^2,0)\right],\\
\Pi_{A,f}^0(p^2) \!\!\!\!\!&=&\!\!\!\!\! \frac{1}{16\pi^2}\frac{m_f^2}{v^2}
\left(-1+\frac{m_{f'}^2}{m_f^2}\right)B_1(p^2,m_{f'}^2,0),\\
\Pi_{S,f}^0(p^2) \!\!\!\!\!&=&\!\!\!\!\! \frac{1}{16\pi^2}\frac{m_f^2}{v^2}
\left(B_0(p^2,m_f^2,M_H^2) - B_0(p^2,m_f^2,0)
-2\frac{m_{f'}^2}{m_f^2}B_0(p^2,m_{f'}^2,0)\right)\,.
\end{eqnarray}
The $z$ and $w$ self-energies only differ in the fermionic part, and are equal
for $m_f=m_{f'}$.  It is easily shown using the explicit results given for
$A_0$ and $B_0$ in Appendix \ref{app.int} that the self energies $\Pi_z^0$ and
$\Pi_w^0$ of the $z$ and $w$ bosons are equal for $p^2=0$, independently of the
values of the masses $m_f$ and $m_{f'}$. This equality does not extend to the
derivatives of the self-energy functions. Finally, comparing the expressions in
Eqs.\ (\ref{tadint}) and (\ref{pizint}), we find that
\begin{equation}
T_0/v_0=\Pi_z^0(0)=\Pi_w^0(0)
\end{equation}
as stated in Eq.\ (\ref{T/v=Pi}).

\section{Loop integral expressions}
\label{app.int}
\subsection{Scalar integral expressions}
\label{sect.scalar}
Here we define the scalar integral expressions needed to calculate the two- and
three-point functions \cite{denner,thooft}. The more complicated vector
and tensor loop integrals which arise from diagrams containing fermions
can be reduced to sums of scalar integrals by standard techniques
\cite{denner,pasvelt}.
We will follow the definitions in \cite{denner}.

For the calculation of the one-loop self-energies in $D$
dimensions we use
\begin{eqnarray}
A_0(m_0^2)&=&\frac{(2\pi\mu)^{(4-D)}}{i\pi^2}\int\! d^Dq\,
                                        \frac{1}{q^2-m_0^2+i\epsilon'}
\label{bint} \,,\\
B_0(p^2,m_0^2,m_1^2)
&=&  \frac{(2\pi\mu)^{(4-D)}}{i\pi^2}\int\! d^Dq\,
\frac{1}{[q^2-m_0^2+i\epsilon'][(q+p)^2-m_1^2+i\epsilon']} \phantom{xxxx}\,.
\end{eqnarray}
The vertex corrections involve the integrals
\begin{eqnarray}
\label{cint}
\lefteqn{C_0(p_1^2,p_2^2,,m_0^2,m_1^2,m_2^2)=} \\
&& \frac{(2\pi\mu)^{(4-D)}}{i\pi^2}\int\! d^Dq\,
\frac{1}{[q^2-m_0^2+i\epsilon'][(q+p_1)^2-m_1^2+i\epsilon'][(q+p_2)^2-m_2^2
+i\epsilon']}\,.\nonumber
\end{eqnarray}
The arbitrary energy scale $\mu$ is introduced to fix the energy dimensions of
the functions $A_0$, $B_0$, and $C_0$ independent of the value of $D$, and the
infinitesimal quantity $i\epsilon'$ defines the integration path in the complex
plane. Note that B\"{o}hm et al. \cite{boehm} define $A_0$ with an overall
minus sign, while Kniehl \cite{kniehl,kniehl2} introduces a minus sign in the
definition of $C_0$.

These integrals can be evaluated in a straightforward manner using Feynman
parameters. Complete analytic results are given by Denner \cite{denner}, though
it is useful to go back to the Feynman representation of the integrals for
certain values of $p_i^2$ and $m_i^2$ to avoid problems with infrared
singularities.

The integral $B_1$ that appears in Eq.\ (\ref{ctfer}) is given by
\begin{equation}
B_1(p^2,m_0^2,m_1^2)=\frac{1}{2p^2}\left[A_0(m_0^2)-A_0(m_1^2)
+(m_1^2-m_0^2-p^2)B_0(p^2,m_0^2,m_1^2)\right].
\end{equation}

The derivatives of $A_0$, $B_0$, and $B_1$ with respect to $p^2$, the square of
the external momentum, are also needed in the calculation of the multiplicative
wavefunction renormalization constants $Z_i$. The function $A_0$ actually does
not depend on the external momentum, and its derivative with respect to $p^2$
therefore vanishes.  For the remaining derivatives we introduce the usual
notation
\begin{equation}
\partial B_i(M^2,m_0^2,m_1^2) \equiv
\left.\frac{\partial}{\partial p^2} B_i(p^2,m_0^2,m_1^2) \right|_{p^2=M^2}\,,
\qquad i=0,1.
\end{equation}
These derivatives are given, for example, by Denner \cite{denner}.

The reduction of tensor integrals such as those which appear in Eqs.\
(\ref{t_i}) and (\ref{t_ii}) to sums of scalar integrals is discussed
in detail in \cite{denner}.The results given there agree with our
calculations \cite{riesselmann} except for a typographical error
in Denner's expression for the function $C_{00}$ in his Eq.\ (C.37).
In our notation, the correct result
for $C_{00}$ is
\begin{equation}
\label{denner.corr}
C_{00}= \frac{1}{4}\left[B_0({(p_1-p_2)^2},m_2^2,m_1^2)
                   + (m_0^2 - m_1^2 + p_1^2)C_1
                   + (m_0^2 - m_2^2 + p_2^2)C_2
                   + 2m_0^2C_0\right].
\end{equation}


%
%
\newpage

\newpage

\newpage

\newpage

\newpage

\newpage

\newpage

\begin{thebibliography}{99}
\bibitem{ReMH}
We follow the convention for an unstable Higgs boson of equating $M_H^2$
with the value of $p^2$ at which the real part of the expression on the
righthand side of Eq.\ (\ref{massshell}) vanishes for $p^2=p^2
+i\epsilon$, $\epsilon\rightarrow 0+$. $\delta m_0^2$, $v_0$,
and $\lambda_0$ are taken as real. With this definition, $\Gamma_{H_0}^{(2)}
\approx Z_H^{-1}(p^2-M_H^2+iM_H\Gamma_H)$ for $p^2\approx M_H^2$,
with $\Gamma_H=-Z_H{\rm\/ Im}\Pi^0_H(M_H^2)$. A definition of $M_H^2$ as the
complex value of $p^2$ at which the expression in Eq.\ (\ref{massshell})
vanishes exactly (i.e., a pole mass in terms of the propagator),
requires some changes which we will not discuss here.
\bibitem{fermionmass}
For an unstable fermion, we will use a definition analogous to that
used for the Higgs boson \cite{ReMH}, and define the square of the
fermion mass as the vale of $p^2$ at which the real part of the
denominator in Eq.\ (\ref{sf}) vanishes. Other definitions are also
possible.
\end{thebibliography}
\end{document}